%% file: main.tex
\documentclass[sigconf]{acmart}

\AtBeginDocument{%
  }

\input{macros}

\copyrightyear{2026}
\acmYear{2026}
\setcopyright{cc}
\setcctype{by}
\acmConference[ICSE-SEIP '26]{2026 IEEE/ACM 48th International Conference on Software Engineering}{April 12--18, 2026}{Rio de Janeiro, Brazil}
\acmBooktitle{2026 IEEE/ACM 48th International Conference on Software Engineering (ICSE-SEIP '26), April 12--18, 2026, Rio de Janeiro, Brazil}
\acmPrice{}
\acmDOI{10.1145/3786583.3786858}
\acmISBN{979-8-4007-2426-8/2026/04}

\begin{document}

\title{Abstain and Validate: A Dual-LLM Policy for Reducing Noise in Agentic Program Repair}

\author{
Jos{\'e} Cambronero$^1$, Michele Tufano$^1$, Sherry Shi$^1$, Renyao Wei$^1$, Grant Uy$^1$, Runxiang Cheng$^1$, Chin-Jung Liu$^1$, Shiying Pan$^1$, Pat Rondon$^1$,
Satish Chandra$^2$
}\authornote{Work done while at Google.}
\affiliation{%
  $^1$\institution{Google, USA}
  \country{}
}
\affiliation{%
  $^2$\institution{Meta, USA}
  \country{}
}

\email{
{
jcambronero, tufanomichele, sherryyshi,
renyaow,
grantuy,
chengsam,
chinjungliu,
shiyingp,
rondon
}@google.com}
\email{
schandra@acm.org
}

\renewcommand{\shortauthors}{Cambronero et al.}

\newif\ifFORARXIV
\FORARXIVtrue

\input{abstract}

\maketitle
\input{intro}
\input{background}

\input{overview}

\input{abstention}

\input{validation}
\input{composing}
\input{evaluation}
\input{results}
\input{discussion}

\input{threats}
\input{related}

\input{conclusion}

\bibliographystyle{ACM-Reference-Format}
\bibliography{references}
\balance

\ifFORARXIV
\clearpage
\appendix
\input{appendix}

\fi

\end{document}

%% file: macros.tex
\usepackage{listings}
\usepackage{xcolor}
\usepackage{booktabs}
\usepackage[symbol]{footmisc}

\newcommand{\boldlead}[1]{\vspace{0.3em}\noindent{\textbf{#1}}}
\newcommand{\leadbold}[1]{\boldlead{#1}}

\definecolor{codegray}{gray}{0.95}

\DeclareMathOperator{\E}{\mathbb{E}}

\usepackage{subcaption} %
\usepackage{xspace}
\usepackage{xcolor}
\usepackage{tabularx}
\usepackage{booktabs}
\usepackage{ragged2e}

\newcommand{\authortodo}[2]{}

\newcommand{\Space}[1]{}
\newcommand{\Comment}[1]{}

\newcommand{\CodeIn}[1]{\begin{small}\texttt{#1}\end{small}}

%% file: abstract.tex
\begin{abstract}
Agentic Automated Program Repair (APR) is increasingly tackling complex, repository-level bugs in industry, but
ultimately these patches still need to be reviewed by a human before committing them to ensure they address the bug.
Showing patches unlikely to be accepted can lead to substantial noise, wasting valuable developer time and eroding trust in automated code changes.
We introduce two complementary LLM-based policies to reduce such noise: bug abstention and patch validation policies.
Bug abstention excludes bugs that the agentic APR system is unlikely to fix. 
Patch validation rejects patches that are unlikely to be a good fix for the given bug. 
We evaluate both policies on three sets of bugs from Google’s codebase, and their candidate patches generated by an internal
agentic APR system.
On a set of 174 human-reported bugs, removing bugs and patches rejected by our policies can raise success rates by up to 13 percentage points and 15 percentage points, respectively, and by up to 39 percentage points in combination.
On null pointer exceptions and sanitizer-reported bugs with machine-generated bug reports, patch validation also improves average single-sample success rates.
This two-policy approach provides a practical path to the reliable, industrial-scale deployment of agentic APR systems.
\end{abstract}

%% file: intro.tex
\section{Introduction}

A substantial portion of software development involves diagnosing and repairing bugs.
To mitigate this cost, the research community has long invested in Automated Program Repair (APR)~\cite{nguyen2013semfix,le2011genprog,long2015staged, le2019automated}.
More recently, advanced language models have enabled more complex, autonomous ``repository-level'' program repair, where a system manages diagnosis, localization, and repair.
Systems like SWE-Agent~\cite{yang2024swe}, AutoCodeRover~\cite{zhang2024autocoderover}, SpecRover~\cite{specrover}, and RepairAgent~\cite{icse2025-RepairAgent} show that agent-based APR can tackle such a challenge, achieving pass@1 bug resolution rates up to 75.20\% on benchmarks like SWE-Bench-Verified (TRAE as of September 2, 2025~\cite{gao2025trae}).

In an industry setting, however, maximizing the rate of successful patches shown to a developer from an APR system is as important as maximizing the system's overall bug resolution rate.
While pass@1 rates are improving, deploying these agents in production still requires a human developer to review and validate the generated patches.
Because developer time is limited, reviewing patches for intractable bugs or unlikely fixes wastes valuable time.
Reducing such noise from unproductive APR attempts is thus crucial, not only to save developer time but also to build developer trust in automated code changes, which is essential for AI tool adoption~\cite{brown2024identifying, noller2022trustenhancementissuesprogram}. 
This work aims to tackle this challenge by proposing two policies.

First, inspired by abstention in classification~\cite{chow2009optimum, herbei2006classification, guo2010characterizing}, we introduce bug abstention (Section~\ref{sec:abstention}).
For human-filed bugs, we use an LLM to predict if our APR system is likely to produce an acceptable patch for a given bug, using only the bug report information.
This check is performed \emph{before} the agent attempts a repair.
If the predicted probability of success is below a threshold, we abstain from running the repair agent.
This avoids showing developers unlikely patches and saves compute costs from futile repair attempts.

Second, for any patches produced, we employ a multi-stage patch validation policy to predict if the patch is correct (Section~\ref{sec:validation}). This policy composes deterministic checks (e.g., running tests) and an LLM-based filter.
The LLM filter first generates a ``fix specification''---a guideline for a correct fix---based on the bug description and original source files. In our evaluation, this is done without knowledge of any ground-truth solution. A second LLM call uses this specification and the agent's trajectory to produce a boolean (fixed or not) outcome, an explanation, and a confidence score derived from token-level log probabilities.

\begin{figure*}[t]
    \centering
    \includegraphics[width=1.5\columnwidth]{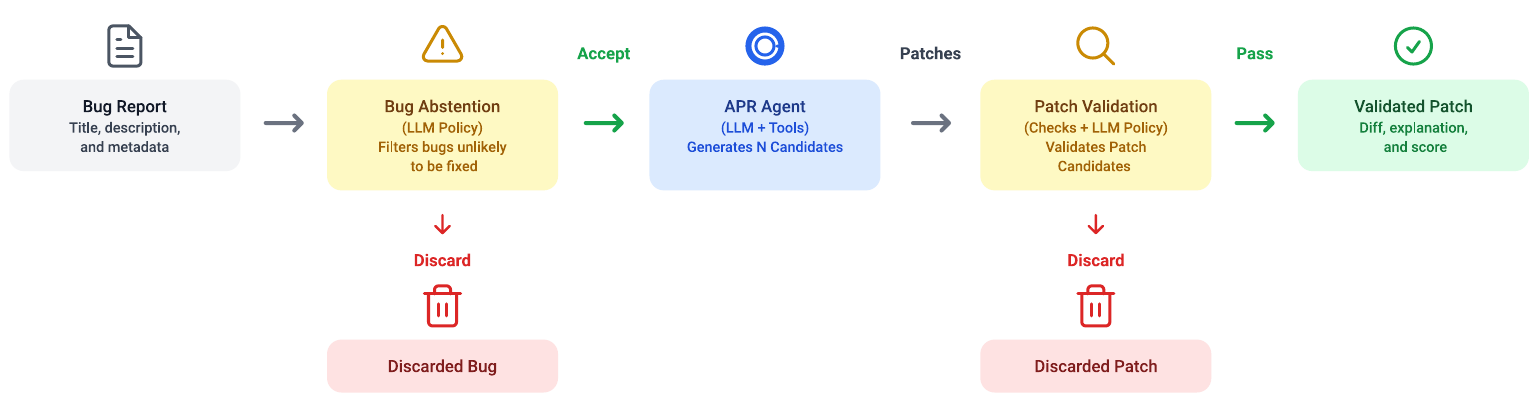}
    \caption{
    Bug abstention rejects issues that are unlikely to be fixed by our APR system. After the agent generates patches, validation decides whether these are likely to have addressed the issue. Jointly they reduce developer noise from APR. 
    }
    \label{fig:lgtm-ensemble}
\end{figure*}

We evaluate the bug abstention and patch validation policies both independently and in combination (Section~\ref{sec:composing}). 
Our combined, two-policy approach first employs bug abstention to screen out bugs unlikely to be fixed, then runs the APR system on the remaining bugs, and finally uses patch validation to filter the generated patches. Our evaluation shows these policies are complementary, improving success rates most effectively when combined. While patch validation has access
to the bug report, we found that a dedicated abstention stage can filter
out bugs where the validation policy would otherwise incorrectly accept a patch.

We evaluate our policies on three sets of bugs from Google's codebase (Section~\ref{sec:evaluation}).
Similar to recent work~\cite{rondon2025evaluating, jimenez2023swe, maddila2025agentic}, the first set has 174 human-reported bugs with known ground-truth fixes and fail-to-pass reproduction tests.
For this set, we evaluate both policies on repair trajectories generated by our ReAct-style APR system.
We further study patch validation on 198 machine-reported Java null pointer exception (NPE) bugs from a live deployment and 50 historical, sanitizer-reported bugs with known reproduction tests.
(We do not study bug abstention on machine-reported bugs as we assume them to be eligible for APR by construction.)

Because our goal is to reduce developer noise, we introduce ``filtered success@k''. This metric computes success (e.g., fail-to-pass or human acceptance) only on the subset of patches and bugs that are ``allowed'' by our policies, rather than over the entire population.

On our human-reported bug set, the baseline fail-to-pass@1 rate is 0.11. Applying bug abstention or patch validation alone raises this rate to 0.21 or 0.29, respectively. When combined, these policies form a configurable funnel that significantly improves the quality of patches shown to developers. A moderately-permissive setting raises the filtered fail-to-pass@1 to 0.35. A stricter funnel configuration raises this rate to 0.53---a 1-in-2 success rate---at the cost of showing patches for fewer bugs. On machine-reported bug sets, patch validation alone also shows significant gains: for NPE bugs, it raises the filtered accept@1 from 0.38 to 0.62, and for sanitizer bugs, it improves the rate by up to 15 percentage points over a baseline using known reproduction tests.

To summarize, our contributions in this work are:
\begin{itemize}
\item A bug abstention policy to select bugs suitable for agentic APR.
\item A multi-stage patch validation policy to assess candidate patch correctness.
\item An extensive evaluation on three industrial bug sets.
\end{itemize}

%% file: background.tex
\section{Background: ReAct-style Agentic APR}
\label{sec:background}

Our agentic APR system~\cite{rondon2025evaluating} consists of an LLM-driven ReAct-style loop with coding tools designed to interact
with an isolated Google environment~\cite{Potvin2016GoogleMonorepo}.
At the start of the process, the agent is provided with a bug report, which may contain varying degrees of information about the underlying bug.
The agent then proceeds for a sequence of up to $N$ steps, where each step is defined as one LLM output and an associated tool execution.
The tools are: code search, viewing a file, editing a file, running tests, and finishing the repair trajectory.
The LLM output consists of two portions: a thought, describing the agent’s intentions in natural language, and an action that is parsed to execute the associated agent tool.
The LLM output and environment responses are added to an append-only history.%

%% file: overview.tex
\section{Abstention and Validation Policies}
\label{sec:overview}

We introduce two key components into an agentic APR system to reduce developer noise. In Figure~\ref{fig:lgtm-ensemble}, we augment the core APR Agent module (in blue) with a component before the repair loop (Bug Abstention) and a component after the repair loop (Patch Validation). The bug abstention policy filters out bugs that the APR agent is unlikely to fix, while the patch validation policy filters out patches that are unlikely to be a correct fix for a given bug.

%% file: abstention.tex
\subsection{Bug Abstention Policy}
\label{sec:abstention}

Abstention is a well-known technique in classification~\cite{cortes2016boosting}, and amounts to allowing a $K$-label classifier to produce an extra label corresponding to “unknown” (i.e. abstaining from making a decision). Analogously, in the repair setting, we say a system abstains from a bug if it decides that it will not attempt repair. 

Importantly, the abstention policy can only access the bug, which comprises textual bug title, description, and accompanying metadata, to make its decision; it does not receive any additional human feedback.
We also choose not to provide codebase access to the abstention policy, which may otherwise require making the policy agentic and losing some efficiency gains.

We choose to model this problem in a probabilistic fashion to produce a score, rather than a boolean decision.
Intuitively, we want this score to approximate the inherent probability of a single repair attempt succeeding. We then convert this into a decision by comparing it to a configurable threshold, $\tau$.

\subsubsection{Formalization}
\label{sec:abstention-formal}

Let $b \in \mathcal{B}$ be a bug in a set of bugs that has been provided to our repair system as they require code changes for their resolution.
We assume that knowing that these bugs require explicit code changes is already determined in a pre-filtering phase or because a user has informed the system. 
Let there be $n$ independent Bernoulli trials reflecting $n$ independent repair attempts by the repair agent on $b$.
The outcome of the $i$-th trial for bug $b$, $X_{b,i}$, is 1 if the agent generates a successful patch in trial $i$ and 0 otherwise.

Accordingly, we define the \textbf{empirical probability of the agent's success on bug $b$} as $\hat{\pi}_{b} = \frac{1}{n}\sum_{i=1}^{n} X_{b, i}$.

This empirical value $\hat{\pi}_{b}$ is an observation-based
estimate of the \textbf{unknown, true probability of success for bug $b$}, with respect to a particular repair system, which we denote as $\pi_b = P(X_{b, i} = 1)$.
Estimating  $\hat{\pi}_b$ requires a success criterion, such as a
test oracle or human reviewer.
During deployment, however, such a criterion is either unavailable (no test) or renders the estimation unnecessary (manual review on all patches).
Thus, for a given bug $b$, we leverage an abstention model $P_a$ to produce $P_a(b)$, which then is used to  estimate $\hat{\pi}_b$, \textit{a priori}.
We aim to turn $P_a(b)$
into a decision policy that allows us to filter down
the original bug set $\mathcal{B}$ to a new set $\mathcal{B'}$, such that 
$\mathcal{B'} \subseteq \mathcal{B}$ and
$\E_{b' \in \mathcal{B}'}[ \pi_{b'}] \geq \E_{b \in \mathcal{B}}[ \pi_b]$, demonstrating a reduced probability
of wasted repair attempts.

Explicitly, the abstention model’s prediction is turned into a decision policy which maps the model's predicted success likelihood to an action from the set $\{\CodeIn{Attempt Repair}, \CodeIn{Abstain} \}$. Given a pre-defined threshold $\tau \in [0,1]$, the policy is to \CodeIn{Attempt Repair} if $P_a(b) \ge \tau$, and \CodeIn{Abstain} otherwise.

Our abstention policy queries an LLM to produce a prediction of the (unobserved) probability of an attempt successfully addressing a given bug. We use an instruction-style prompt and instruct the model to produce one of two possible outcomes, \CodeIn{success} (attempt repair) or \CodeIn{failure} (abstain), given a prompt with some representation $r(b)$ of the given bug $b$. We define $P_a(\CodeIn{outcome}|r(b))$ as the model-predicted probability of that token. Since \CodeIn{success} and \CodeIn{failure} are the only two possible outputs based on the instructions, the LLM’s decoding should place relatively little weight on other tokens and so $P_a(\CodeIn{success}|r(b))+P_a(\CodeIn{failure}|r(b)) \approx 1$. We confirmed this property on a held-out set of bugs during our development. We note that the raw LLM probabilities are not necessarily calibrated to the true empirical success rates $\hat{\pi}_b$. However, strict calibration is not essential for abstention performance, provided the model assigns higher scores to bugs that are more likely to be resolved.

\subsubsection{Bug representations}
\label{sec:abstention-bug-representations}

As is well known, the prompt used to query an LLM can have a substantial impact on its performance~\cite{lu2024prompts, tafreshipour2025prompting}. We describe two possible bug representations for our prompt.

\textbf{Instruction Prompting without Guidelines: $r_1(b)$.}~~We included the bug title and description, the capabilities of the agent (i.e., information on the commands available to the agent, execution budget for the expected agent runs) in the prompt. The prompt
instructs the model to analyze the bug's title and description and output only \CodeIn{success} or \CodeIn{failure}. %

\textbf{Instruction Prompting with Guidelines: $r_2(b)$.}~~We built on top of $r_1(b)$ by adding guidelines that can be used to determine the likelihood of successful bug resolution.
We derived these guidelines
through a manual process on a held-out set of 50 human-filed bugs. Four authors inspected the bug titles and descriptions, determined subjective difficulty levels, and provided short guidelines that would help identify bugs that could be resolved by reasonable attempts from an agent. One author collated the outcomes into 5 guideline criteria---``Clear Problem \& Action'', ``Precise Code Localization'', ``Simple, Localized Fix Anticipated'', ``Actionable Reproduction \& Sufficient Context'', and ``Self-Contained and Complete''---with one-sentence definitions of each criterion.

%% file: validation.tex
\subsection{Patch Validation Policy}
\label{sec:validation}

Patch validation performs an analogous accept/reject decision at the patch level. Specifically, it predicts whether a \textbf{particular} repair trajectory produced a successful patch for a specific bug.

\subsubsection{Formalization}

Let $t$ be a particular trajectory from the repair agent when attempting to repair bug $b$. Given some representation of the trajectory, $r(t)$, the policy predicts a triplet $P_v(b, r(t)) \in \{\CodeIn{false},\CodeIn{true}\} \times \Sigma^* \times \mathbb{R}$,
where the first element is a binary judgment for 
patch correctness,  the second element is a natural language explanation of the judgment, and the third element is an optional confidence score for the patch's correctness. We then define a decision function over this triplet, which can use the binary judgment or the patch correctness confidence score to filter out patches and allow for the intelligent allocation of human effort, filtering out less-likely patches as noise and reducing review time.

In our implementation, we compute $P_v$ using a multi-stage approach. We first perform build and test regression checks to remove patches that have known issues. Specifically, we use Google’s unified build system to extract targets that have any modified files as a direct dependency. 
Alternatively, if the bug we are tackling has a known reproduction test, we replace this regression check with a filter that rejects any patches that fail to pass the reproduction test.

Next, our reviewer applies a simple heuristic which removes trajectories that have a final test command issued by the agent that results in a failure. Note that this may capture different test targets than the build/test check as the agent is not restricted to executing only direct dependencies. 
Finally, our validation employs a two-staged LLM prediction, which we now describe in detail.

\subsubsection{Fix specification}

Past literature in the area of LLM-as-judge has explored the use of generated rubrics to perform more accurate scoring of an answer~\cite{tong-zhang-2024-codejudge}. However, this rubric is often based on additional information such as the reference answer~\cite{agentjudge}, which is naturally unavailable in our setting, where we are attempting to repair a program without knowing the fix ahead of time.

However, we do take inspiration from this line of research, and explore the LLM’s ability to generate a set of criteria for determining if an agent attempt has successfully addressed a bug \emph{without} access to any ground truth—we refer to this as a ``fix specification.''
We simply provide this specification as part of the LLM judge prompt to produce a score for the patch. 
To generate the fix specification, we prompt an LLM with the bug information and the \emph{original} source code contents of the files that were edited in the agent’s candidate patch. Importantly, this connects the fix specification and the agent’s attempt by re-using the localization information from the agent’s trajectory to generate the fix specification. Note that this agent trajectory \textit{does not} contain any information from the ground-truth fix—that is, the specification can only be derived from information known at the start of the repair attempt. We do not
impose any format or structure on the
generated specification and instead
allow the LLM to freely generate
this content. In practice, the model often generates clear Markdown with a diagnosis and list of expected properties of a correct change.

\subsubsection{Reviewing with a specification}
\label{sec:validation-with-specification}
Once the LLM has generated a fix specification, a second LLM call judges the agent’s attempt. This prompt includes the bug information, the unified diff--format patch produced from the agent's trajectory, test result information (extracted from the trajectory) for any test targets run by the agent, and the generated fix specification. Note that reviewing with a specification carries the risk of relying on
LLM-generated criteria, which itself may be wrong, and mislead the policy.

A test result consists of a numbered test attempt for a particular test suite target, and the associated environment response (including logs for a failure). If no test result information is found, we simply state ``No testing information is available.''

\subsubsection{Reviewing without a specification}
\label{sec:validation-without-specification}

Besides our specification-based validation, we experiment with a variant of validation that consumes the bug description and the concatenation of all prompt messages and responses that were produced during the agent’s execution. Recent work~\cite{pan2025training} has used this representation to judge an agent’s attempts to solve SWE-Bench tasks.

\subsubsection{Patch validation outputs}
\label{sec:validation-outputs}
As described previously, the output of the validation policy is a triplet consisting of a judgment, a natural-language explanation, and a real-valued confidence score. To enforce this format, we employ constrained decoding~\cite{deutsch2019general, scholak-etal-2021-picard}.

We define the patch correctness confidence score for the response heuristically. If validation returns a negative binary judgment, this score is set to zero. If the validation returns a positive binary judgment, we set this score to the exponential of the average token log probabilities. While not a well-defined probability, our evaluation shows this score can improve performance over binary labels.

\subsubsection{Decisions over patch validation outputs}
\label{sec:validation-decisions}
We can make decisions based on the binary judgment or the patch correctness confidence score extracted from the validation outputs.
We consider two variants: (1) \textbf{binary}, where we use the binary judgment and accept patches with a positive label; (2) \textbf{percentile of patch correctness confidence score}, where we estimate a particular percentile cutoff across all patches for all bugs in the given set (e.g. after applying deterministic filters) and then accept patches with a score at or above that cutoff. While percentile thresholding needs a set of patches to perform a judgment, it mitigates the risks of score drift, e.g. due to population or model changes, and closely aligns with the idea of budget-based reviewing (i.e., a human reviews the top $\le k$ patches).

%% file: composing.tex
\section{Composing Abstention and Validation}
\label{sec:composing}

While abstention and validation policies can be used independently, these two policies can also be combined.  We present and evaluate a simple sequential combination of the two policies. 

Let $R$ represent our repair system, which, given a bug $b$, generates multiple trajectories independently. Each trajectory $t$ has a $\CodeIn{.patch}$ property representing all changes made by the agent. Let $P_a(b)$ be the probability predicted by the abstention policy for sampling a successful fix for bug $b$. Let $P_v(b,t)$ be the patch validation policy output. Let $\tau_a$ be the abstention threshold and $D_v$ be the decision function over the validation’s output. Then the set of patches ultimately shown for human review is: $\{t.\CodeIn{patch} | P_a(b) \ge \tau_a, t \in R(b), D_v(P_v(b, t))\}$. We leave alternate compositions (e.g., probability based) or further optimizations (e.g., accounting for bug/review cost and fix value) for future work.

%% file: evaluation.tex
\section{Evaluation}
\label{sec:evaluation}

\input{tables/dataset}

We now describe our datasets, metrics, and research questions.

\subsection{Datasets}

We use three industrial bug datasets (Table~\ref{tab:data-and-metrics}) for evaluation.
Bug abstention only applies to human-reported bugs.
Machine-reported bugs (i.e., NPE and sanitizer bugs) are all good candidates for APR by construction, as they contain reproduction tests or stack traces.

\subsubsection{Human-reported bugs:}

From Google’s monorepo, we collected a set of 174 human-reported bugs filed within a year by following a curation process similar to that for popular benchmarks like SWE-Bench and recent agentic APR work in industry~\cite{rondon2025evaluating, maddila2025agentic}. Specifically, we adhere
to the curation process detailed in ~\cite{rondon2025evaluating} and additionally
restrict examples to those with at most $31$ 
test targets that depend on the associated ground truth files.

We then generated 20 trajectories 
with a 25-step limit
for each of the 174 bugs using our agentic APR system (Section~\ref{sec:background})
using Gemini 2.5 Pro~\cite{gemini2p5} with temperature 0.2 and top\_p 0.95.

We consider the success criterion to be whether the held-out fail-to-pass test succeeds. We have found this criterion closely matches correctness in human-reported bugs from our codebase~\cite{rondon2025evaluating}.

\subsubsection{Null pointer exception (NPE) bugs}
\label{sec:eval:dataset:npe}
We collected a set of 198 machine-reported NPE bugs that our agent attempted to repair during a live deployment. Every bug report includes a stack trace of the exception. The report may also include a section (\CodeIn{message blob}) that provides additional machine-generated debugging information.

During deployment, our system used patch validation with fix specification to filter patches, normalize patches, and choose a single patch per bug based on majority voting. 
The normalization step includes normalizing identifier strings (mapping each unique likely identifier string to a unique index) and post-patch least-line-count tie-breakers.
A patch was then sent for human review, and submitted to the codebase if a code owner accepted it as fix.

During live deployment, our APR setup differed in the following:
function calling is through Gemini instead of custom ReAct parsing, asynchronous step execution allows long-running calls without blocking concurrent repairs nor overloading services, and at most 5 trajectories were generated with a 25-step limit per bug.

\subsubsection{Sanitizer bugs}
We collected a set of 50 sanitizer-reported bugs with a known ground-truth fix following the curation procedure described in 
our past work~\cite{rondon2025evaluating}. These bugs cover issues such as memory errors, data races, datatype overflows, typecasting errors, and fuzzer-detected errors. We use the same APR agent configuration used for our human-reported bug set.

\subsection{Metrics}
We report precision and recall for both policies. For repair performance, we focus on a standard pass@k metric, and a variant where bugs and patches are filtered by the corresponding policies.
Let $\mathcal{B'}$ be the set of bugs with at least one repair run/trajectory available after 
bug abstention has filtered bugs and patch
validation has filtered patches,
$n'$ be the
total number of runs remaining after applying
patch validation, and $c'$ be the successful count among those. Then the filtered success@k metric is defined as:
$$
\text{filtered success}@k=\mathbb{E}_{\mathcal{B'}}[1 - \frac{\binom{n' - c'}{k}}{\binom{n'}{k}}]
$$
The success criterion differs by bug types.
For human-reported bugs, we compute fail-to-pass on the held-out ground-truth test and report filtered fail-to-pass@k. For NPE bugs, which were open at the time of collection and may not have any tests, we use human acceptance annotations and report filtered accept@k. 
The annotations are from real code owners, or authors if the patch had not been reviewed by owners during live deployment, e.g., a patch not picked by majority voting (\S\ref{sec:eval:dataset:npe}).
For sanitizer bugs, which have known ground-truth fixes and a reproduction test at reporting time for filtering patches, we manually compare patches to ground-truth fixes for semantic equivalence, and report filtered accept@k.

For both policies, we also include comparisons to a no-skill policy which always accepts bugs (or patches, respectively). Under this no-skill policy, the precision of predictions corresponds to the prevalence of successful cases in the population.

\subsection{Research Questions}

We aim to answer the following four research questions:

\leadbold{RQ1}: Can bug abstention successfully abstain on our human-reported bugs? Does this improve filtered fail-to-pass@k?

\leadbold{RQ2}: Can patch validation successfully predict if a patch should be rejected on our human-reported bugs? Does this improve filtered fail-to-pass@k?

\leadbold{RQ3}: Does combining bug
abstention and patch validation
further improve filtered fail-to-pass@k on our human-reported bugs?

\leadbold{RQ4}: How does patch validation improve filtered accept@k on bugs with systematically-generated reports (NPE and sanitizer bugs) and known reproduction tests (sanitizer bugs)?

%% file: tables/dataset.tex
\begin{table}[]
  \centering
  \resizebox{\columnwidth}{!}{
\begin{tabular}{p{0.3\columnwidth} p{0.04\columnwidth} p{0.31\columnwidth} p{0.35\columnwidth}}
    \toprule
    \textbf{Benchmark} & \textbf{\#} & \textbf{Policies} & \textbf{Success \mbox{Criterion}} \\
    \midrule
    \textbf{Human}-reported bugs with real fix. & 174 & Abstention+Validation & Passes held-out reproduction test. \\ \hline
    \textbf{NPE} bugs from a live deployment of a repair system. & 198 & Validation & Confirmed correct by code owner or manual inspection. \\ \hline
    \textbf{Sanitizer}-reported bugs with real fix. & 50 & Validation & Confirmed correct by manual inspection. \\ 
    \bottomrule
\end{tabular}}
  \label{tab:datasets_compact} %
\caption{Benchmarks, policies applied, and success criteria.}
\label{tab:data-and-metrics}
\vspace{-10pt}
\end{table}

%% file: results.tex
\section{Results}

We now present our experimental results.
At the time of these experiments, our abstention policy uses Gemini 2.0 Flash (newer models are now available), our patch validation policy uses Gemini 2.5 Pro, and both perform greedy decoding (temperature = 0). 
We expect to use a smaller model for abstention for efficiency as it runs on a larger number of inputs than patch validation.

\subsection{RQ1: Bug Abstention Policy}

We evaluate bug abstention on the two bug representation variants: with and without the use of general guidelines (Section~\ref{sec:abstention-bug-representations}).

\begin{figure}
\includegraphics[width=0.8\columnwidth]{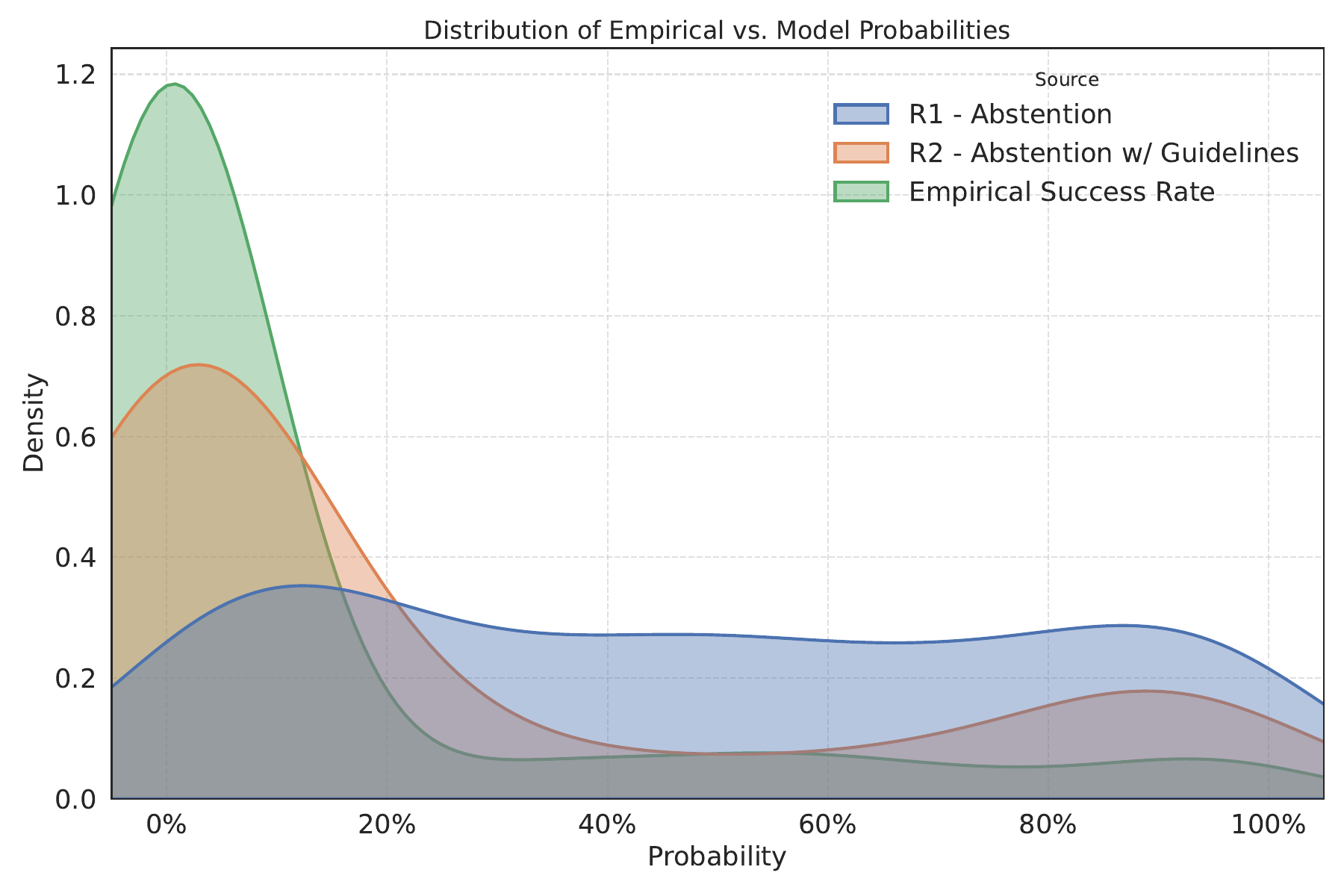}
\caption{
Using abstention with guidelines better aligns model probabilities with observed (left-skewed) rates.}
\label{fig:rq1-pdf}
\end{figure}

Figure~\ref{fig:rq1-pdf} shows a comparison of model-predicted probabilities and the empirical success rate (Section~\ref{sec:abstention-formal}) of repair runs for the 174 human-reported bugs.
The empirical success rate (green) is the fraction of trajectories that have a passing outcome on the held-out reproduction test, which indicates the inherent difficulty of APR.
As shown, it is heavily skewed, with the vast majority of bugs having a very low true success rate, concentrated in the 0--15\% range. 
The skewness confirms that an agentic APR system can benefit from an abstention mechanism to avoid wasting human review effort.

Abstention with representation $r_1$ (i.e., without guidelines) produces a more uniform distribution of repair success probability. 
In contrast, abstention with $r_2$ (i.e., with guidelines) produces a distribution that aligns much more to that of the empirical success rate. 
We further measured the Wasserstein distance~\cite{arjovsky2017wasserstein} between each predicted distribution and the empirical success rate distribution: the distance from $r_2$ (0.126) was nearly three times smaller than that from $r_1$ (0.363), indicating that $r_2$ is a much better overall fit for bug abstention.
Like the empirical data, abstention predictions with $r_2$ are correctly concentrated at lower probability values, while still assigning a higher repair success probability to easier bugs.

\begin{figure}
\includegraphics[width=0.75\columnwidth]{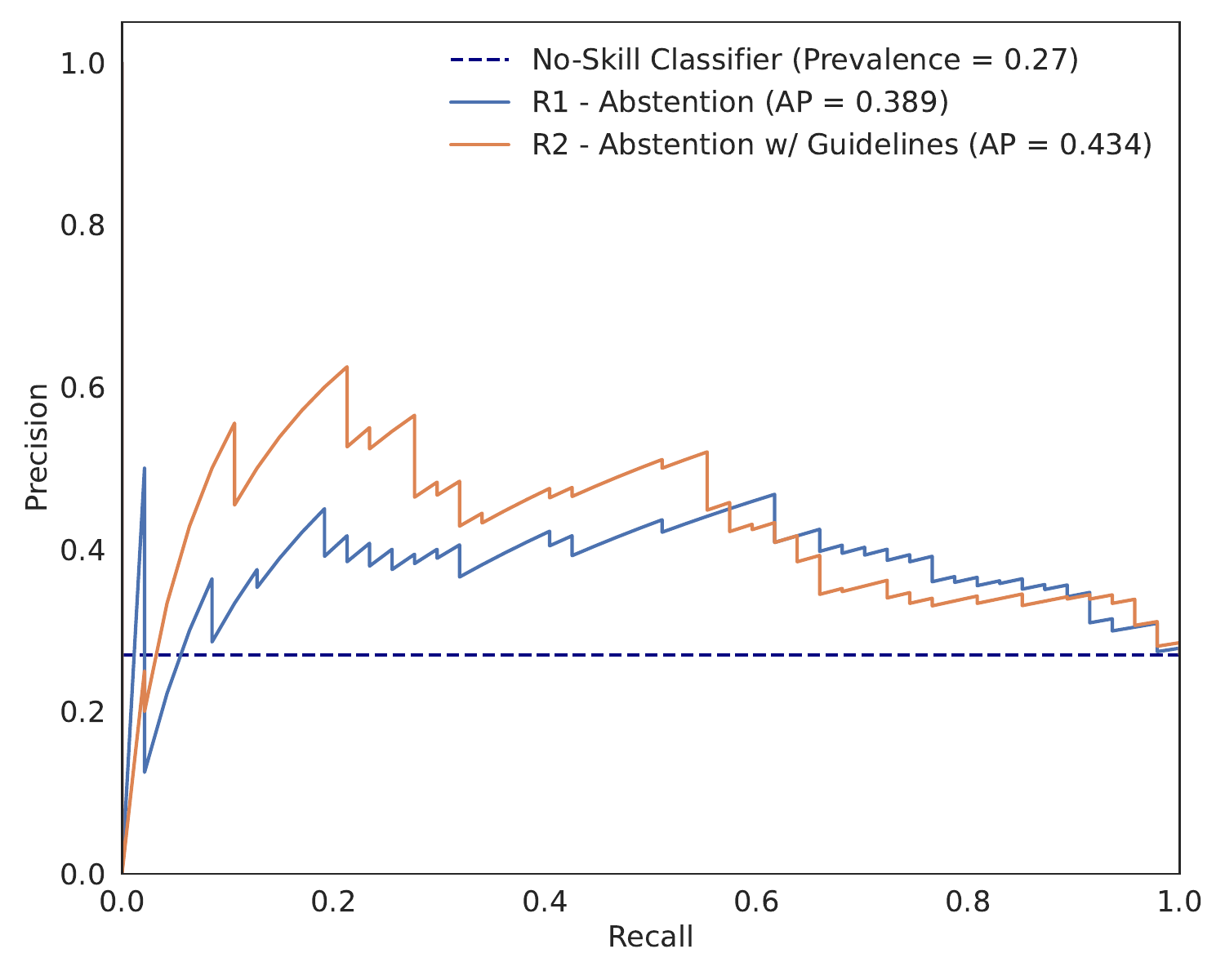}
\caption{Both abstention variants outperform randomly sampling to identify bugs with 
at least one successful repair. Adding guidelines helps. (AP = average precision)
}
\label{fig:rq1-pr}
\end{figure}

Figure~\ref{fig:rq1-pr} presents precision-recall curves for both abstention policy variants, a horizontal line for the prevalence of a success, and a summary average precision (AP) score.
For purposes of this classification, we define success as any bug that has at least one repair trajectory that passes the held-out reproduction test among the 20 trajectories of the bug.
The prevalence of success corresponds to the precision that could be expected from a no-skill policy, which just accepts all bugs, and all patches for these bugs.  
For our human-reported bug set, this prevalence corresponds to 27\% of all bugs having at least one successful repair trajectory without any abstention filtering.
We observe that both abstention policy variants can improve over prevalence; adding guidelines further provides improvements across a range of recall values.

To turn the abstention predictions into a decision policy, we select $\tau=0.5$ as a sensible default for the abstention policy threshold based on a separate development bug set.
Using this threshold produces a precision/recall of 0.44/0.41 in that set.
This threshold is also intuitive, as it prioritizes bugs with a predicted success probability greater than failure.
More importantly, it achieves a strong balance between precision and recall, effectively filtering out a large number of bugs unlikely to be fixed while retaining a significant portion of the fixable ones.

\begin{figure}
\includegraphics[width=0.75\columnwidth]{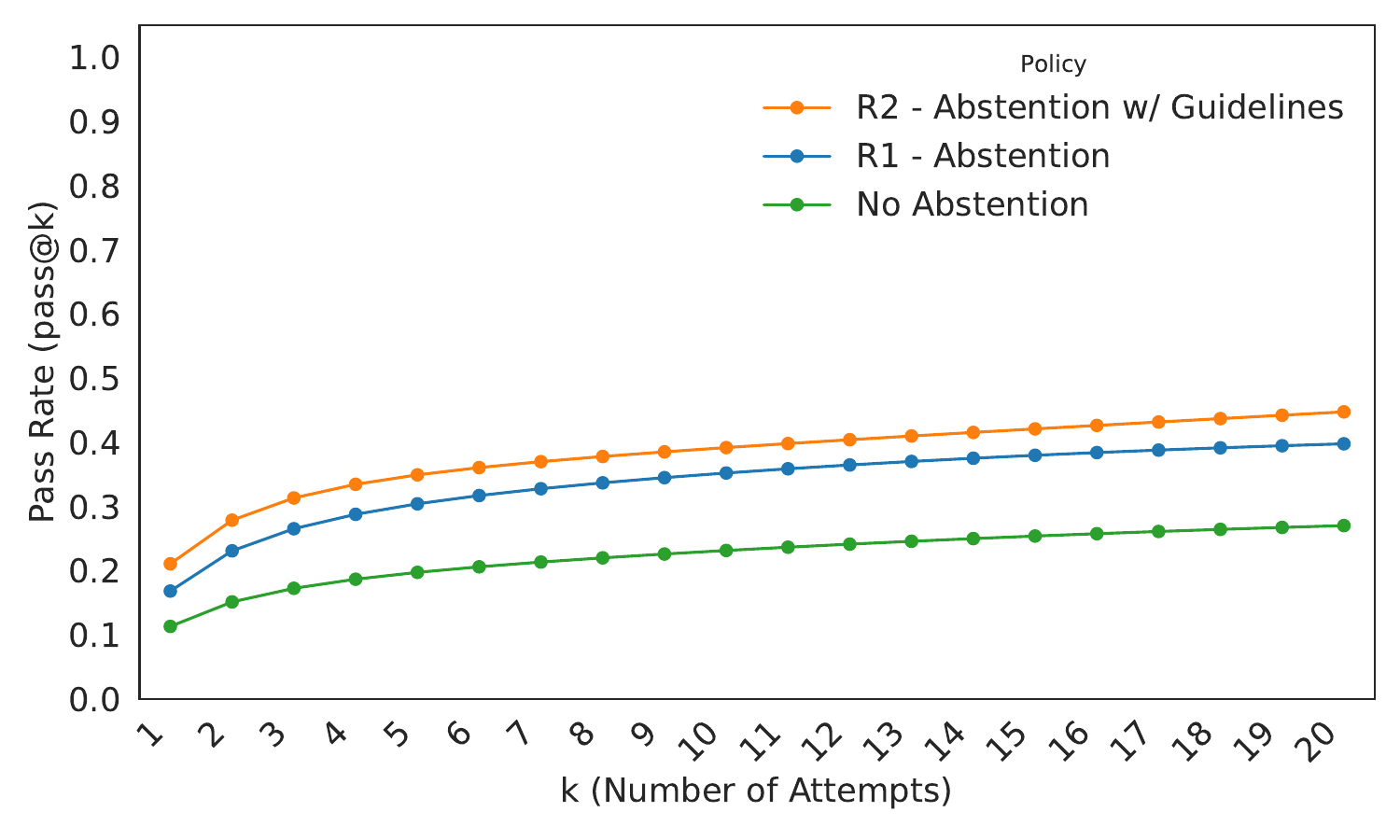}
\caption{Abstention can raise filtered fail-to-pass@k by removing bugs that are less likely to succeed with our system.}
\label{fig:rq1-pass-at-k}
\end{figure}

Figure~\ref{fig:rq1-pass-at-k} shows the pass@k (when abstention is not applied) and the corresponding filtered fail-to-pass@k (when abstention is applied) of our agentic APR system on the set of 174 human-reported bugs.
We use $\tau=0.5$ when applying abstention.
Applying abstention improves the system's efficacy, and the $r_2$ abstention variant demonstrates the most significant gain.
The baseline (no abstention) achieves pass@1 of 11.29\%. 
By filtering bugs with the $r_1$ abstention policy, the filtered fail-to-pass@1 rate increases to 16.81\%. 
The $r_2$ abstention achieves an even higher filtered fail-to-pass@1 of 21.05\%.
This trend of improvement continues as more repair attempts are made: when $k=20$, (filtered fail-to-)pass@k climbs from 27.01\% (no abstention) to 44.74\% ($r_2$ abstention).
We define a bug as ``fixable'' (ground-truth positive) if the agent produces at least one passing trajectory within 20 attempts. Under this definition, our policy yields a 16.5\% false positive rate, meaning it incorrectly attempts repair on only 16.5\% of the unfixable bugs.
These results strongly suggest that abstention effectively discards bugs with a low fix probability, allowing the APR system to concentrate on more-promising bugs.

\subsection{RQ2: Patch Validation Policy}

We compare three patch validation policy variants: (1) \textit{no-skill policy}, i.e., prevalence, which reports the fraction of successful trajectories after filtering patches that fail build and test regression checks, (2) \textit{validation} that uses fix specification (Section~\ref{sec:validation-with-specification}), and (3) \textit{no-spec-validation} that does not use fix specification and reviews the agent’s whole trajectory (Section~\ref{sec:validation-without-specification}).

To determine the build and test regression checks, we extract from Google’s unified build system any targets that have a modified file as a direct dependency.  We retrieve this direct dependency by identifying the build system rules that include the modified file in their definitions. Given Google’s monorepo, transitive dependencies can quickly grow, and so we use this restricted set of rules as a practical approximation to identifying and removing regressions.

Our experiments show that while using fix specification improves both precision and recall over the ablation, the absolute precision remains modest at 0.3.
The no-spec validation’s binary judgment is heavily biased towards positive judgment, providing low recall of failing patches and prevalence-level precision for successful patches.
The no-spec validation’s bias towards positive judgment results in a high false positive rate (ratio of false positives to ground-truth negatives) of 0.64 compared to 0.28 for the variant that uses the fix specification.

\begin{figure}
\includegraphics[width=0.75\columnwidth]{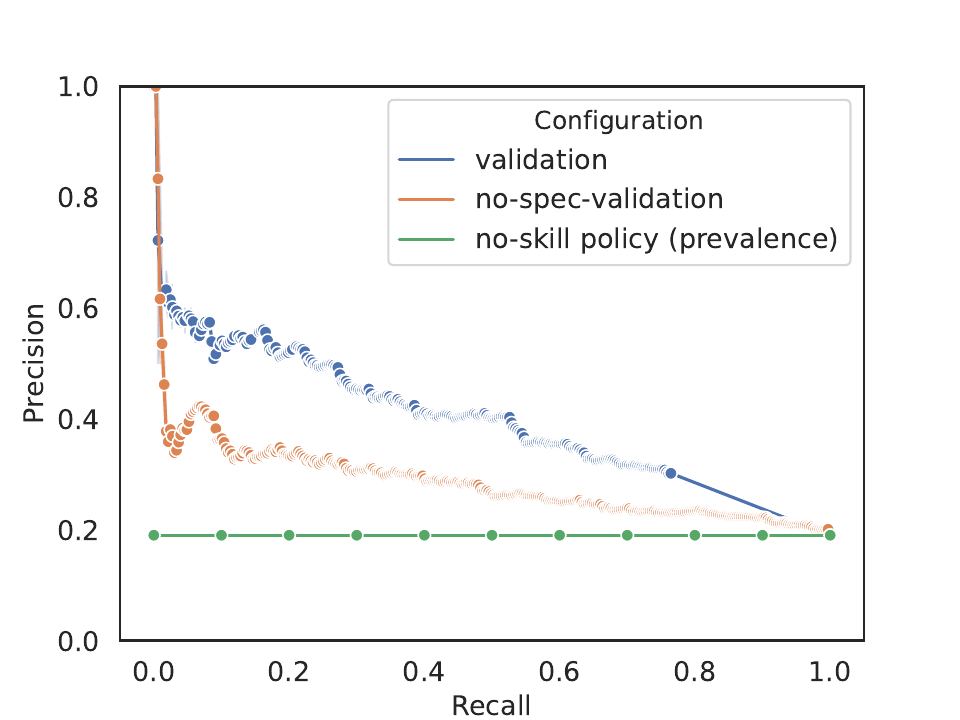}
\caption{Patch correctness score allows more aggressive validation, with precision up to 3$\times$ the success prevalence. The discontinuity for \emph{validation} at high recall is due to a higher fraction of zero correctness
scores (defined in Section~\ref{sec:validation-outputs}).
}
\label{fig:rq2-pr}
\end{figure}

We show that despite the modest predictive power of binary judgment, the validation’s patch correctness score can be used to distinguish between predictions. 
As a reminder, we define this score as zero for negative binary judgments and as the exponential of the average log probabilities of the 
response tokens for positive binary judgments. 
Figure~\ref{fig:rq2-pr} shows precision-recall curves for both the validation and the no-spec validation, as well as a horizontal line for the no-skill policy after regression checks.
At stricter thresholds, our spec-based validation (in blue) can yield a precision up to three times higher (at the cost of recall) than that of the prevalence rate, offering a substantial noise reduction for a subset of bugs.
we find that the no-spec validation can improve over the no-skill policy but achieves lower precision at the same recall than the spec-based validation. 
With the 75th and 90th percentile cutoffs, the spec-based validation achieves a false positive rate (FPR) of 0.13 and 0.04, respectively, while the no-spec validation achieves 0.15 and 0.06.

\begin{figure}
\includegraphics[width=0.8\columnwidth]{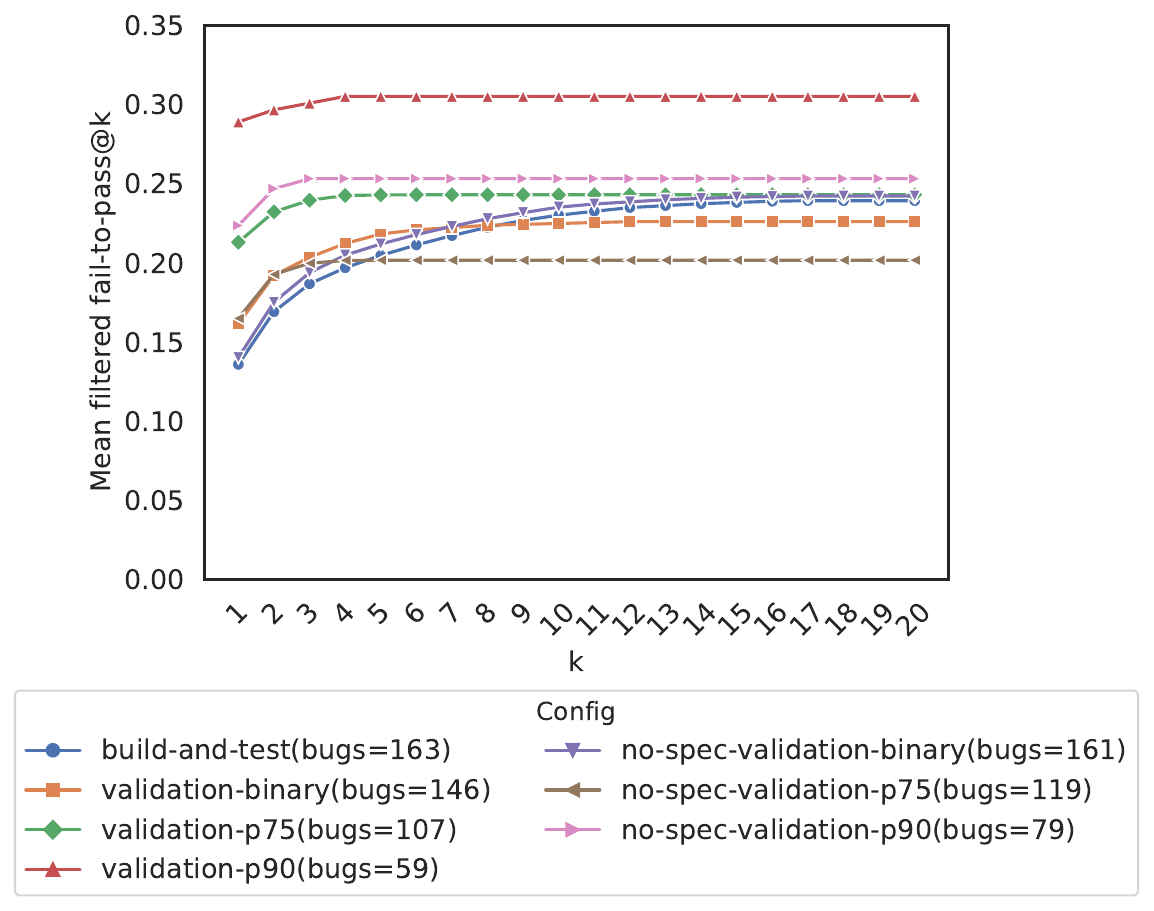}
    \caption{Validation alone can raise filtered fail-to-pass@1 up to 15 percentage points compared to using only build and test regression checks. Parentheses show number of bugs with at least one patch after validation.
    }
    \label{fig:rq2-pass-at-k}
\vspace{-10pt}
\end{figure}

We now study the impact of different decision procedures for patch validation with filtered fail-to-pass@k.
We evaluate two decision procedures described in Section~\ref{sec:validation-decisions} for validation.
Specifically, one procedure uses binary judgment:
\CodeIn{validation-binary} and \CodeIn{no-spec-validation-binary}.
The other uses the 75th and 90th percentile thresholds of all patches’ (after applying deterministic filters) correctness scores: \CodeIn{validation-p75} and \CodeIn{validation-p90} (and similarly for \CodeIn{no-spec-validation}).
Figure~\ref{fig:rq2-pass-at-k} compares the filtered fail-to-pass@k of these patch validation variants against \texttt{build-and-test}, which is a baseline that samples patches that pass build and test regression checks. 
The filtered fail-to-pass@k of \CodeIn{build-and-test} are 0.14 and 0.19 when $k$ equals to 1 and 3, respectively.
On no-spec validation, using binary judgment provides no improvement.
However, using the 75th and 90th percentile thresholds of patch correctness scores, no-spec validation raises filtered fail-to-pass@1 to 0.16 and 0.22, respectively.
Similarly, on spec-based validation, using 75th and 90th percentile thresholds of patch correctness scores better improve filtered fail-to-pass@1 than using binary judgment.
Binary judgment only modestly improves filtered fail-to-pass@1 (0.16), while the 75th and 90th percentile thresholds improve filtered fail-to-pass@1 more substantially to 0.21 (+7 percentage points, +50\% relative improvement) and 0.29 (+15 percentage points, +100\% relative improvement), respectively.
Overall, using an aggressive validation policy results in fewer bugs that have at least one patch that can be surfaced. Given our goal of reducing developer noise, we prioritize increasing success likelihood for remaining patches.

\begin{figure}
\includegraphics[width=0.65\columnwidth]{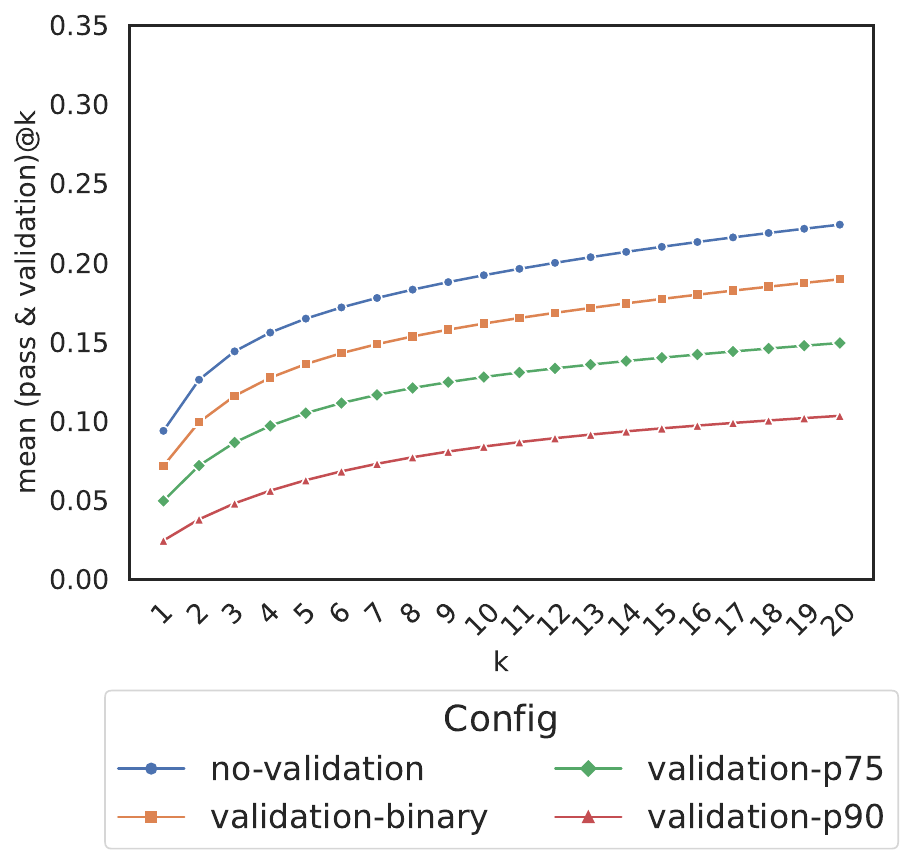}
\caption{More aggressive filtering tradeoffs between reducing developer noise and increasing bug resolution rate.}
\label{fig:rq2-pass-and-critic-success}
\end{figure}

In Figure~\ref{fig:rq2-pass-at-k}, filtered fail-to-pass@k improves further as a validation variant with more aggressive filtering is applied.
While this improvement trend matches our goal on reducing developer noise, it also presents a tradeoff occurring in patch validation between reducing developer noise (by discarding more agent trajectories) and increasing bug resolution rate.

To better understand this tradeoff, we compute ``(pass \& validation)@k'', where at least one patch must both pass the reproduction test and satisfy the validation, and present the results in Figure~\ref{fig:rq2-pass-and-critic-success}.
Figure~\ref{fig:rq2-pass-and-critic-success} shows that: more aggressive filtering in validation reduces overall bug resolution rate.
For example, (pass \& validation)@1 decreases by 2, 4, and 7 percentage points for binary judgment, 75th percentile threshold judgment, and 90th percentile threshold judgment, respectively.
In summary, when employing a validation policy, there is this tradeoff between wasted compute and developer attention to consider.
The validation policy ultimately rejects many generated patches, which lowers overall bug resolution rates and leads to early flattening in Figure~\ref{fig:rq2-pass-at-k}, but ensures the patches shown to developers are more likely to be plausible.

\subsection{RQ3: Combined Abstention-and-Validation}

\begin{figure}
\includegraphics[width=0.7\columnwidth]{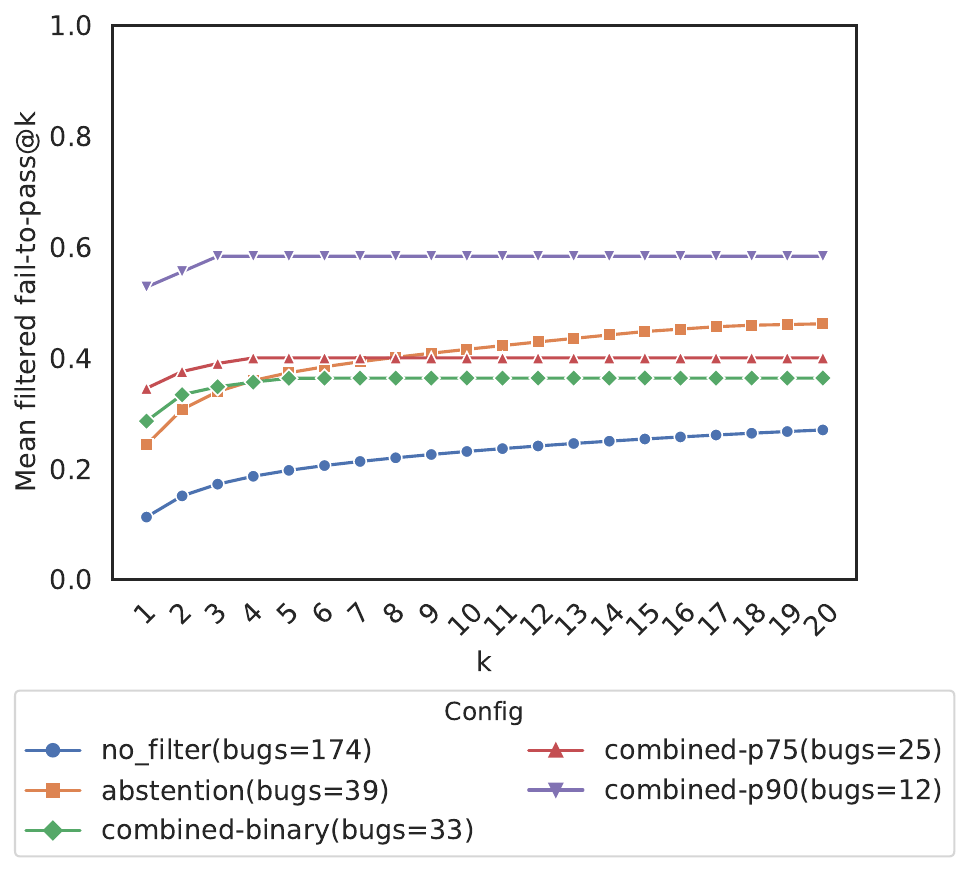}
\caption{Abstention and validation are complementary, jointly they can raise filtered fail-to-pass@k.}
\label{fig:rq3-pass-at-k}
\end{figure}

We now evaluate a dual-policy framework that combines abstention and validation: it filters bugs via the abstention policy, then filters patches via the validation policy (Section~\ref{sec:composing}).
For evaluation, we set the abstention threshold to 0.5 based on experiments on a separate dev set to balance tradeoffs, rejecting any bugs with a predicted value below that. We use spec-based validation, and consider both binary and percentile-threshold judgments.
Figure~\ref{fig:rq3-pass-at-k} compares filtered fail-to-pass@k between the combined variants (\CodeIn{combined-*}), a policy that only applies abstention and excludes trajectories that produce no patches (\CodeIn{abstention}), and a baseline that applies neither abstention and validation (\CodeIn{no-filter}).

Combined abstention and binary validation can raise filtered fail-to-pass@1 to 0.29 over abstention alone (+5 percentage points) or validation alone (+13 percentage points).
More aggressive filtering, such as 75th percentile can increase filtered fail-to-pass@1 to 0.35 (+11 percentage points relative to abstention alone and +14 to the same validation alone).
These percentile thresholds create a configurable funnel for developer review. A strict policy (\CodeIn{combined-p90}) narrows the intake to just 12 bugs, but its output is high-confidence, with a filtered fail-to-pass@1 of 0.53 (a 1-in-2 success rate). A more permissive policy (\CodeIn{combined-p75}) widens the intake, doubling the number of bugs for inspection while still yielding a high success rate of roughly 1-in-3.

\begin{figure}
\includegraphics[width=0.7\columnwidth]{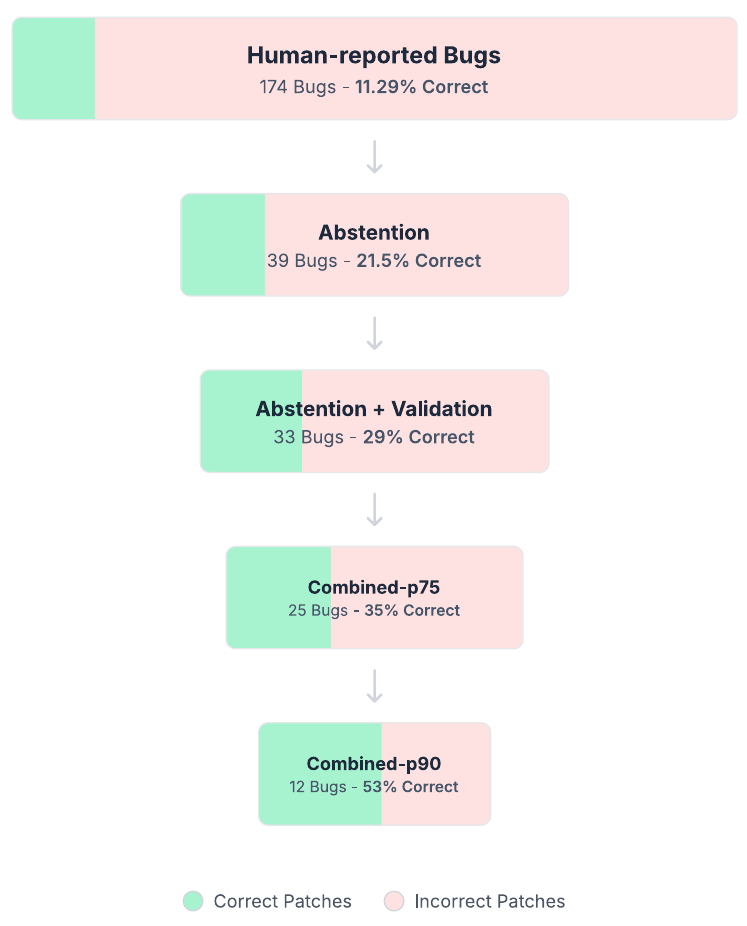}
\caption{
Our combined policies filter an initial set of 174 bugs. Each stage reduces the number of bugs but increases the filtered fail-to-pass@1 rate, from a baseline of 11.29\% to 53\% for the most selective configuration.}
\label{fig:funnel}
\end{figure}

While it is not surprising that using validation can improve on bug-abstention-only success rates, as validation has access to more information (e.g., the agent-generated patch) and is judging a single instance, we also observe improvements over the validation-only version. This is interesting because patch validation---just like the bug abstention policy---has access to the bug report. We hypothesize that, given a patch, validation is more likely to be persuaded of the success of the attempt, while the abstention policy is more likely to reflect overall quality of bug report.

Figure~\ref{fig:funnel} visualizes the process of our dual-policy framework as a configurable funnel.
The funnel presents the filtered fail-to-pass@1 rate on human-reported bugs at each stage. 
Starting with the full set of 174 bugs, where our APR system has a baseline pass@1 rate of 11.29\%, the \CodeIn{abstention} policy first filters this down to 39 more-promising bugs, increasing the filtered fail-to-pass@1 rate to 21.5\%. The combined validation policies (\CodeIn{combined-*}) further refine this set. The \CodeIn{combined-binary} policy leaves 33 bugs with a rate of 29\%. More aggressive filtering with \CodeIn{combined-p75} and the strictest \CodeIn{combined-p90} variant narrow the funnel to 25 and 12 bugs, respectively, boosting the quality of patches to achieve high filtered fail-to-pass@1 rates of 35\% and 53\%. Overall, Figure~\ref{fig:funnel} demonstrates the trade-off between the volume of bugs attempted and the precision of the patches ultimately shown to developers.

\subsection{RQ4: Validation on NPE and sanitizer bugs}

We further evaluate patch validation in bug categories that have more systematic information in their bug reports, and so are assumed good candidates for APR by construction (i.e., these bugs pass abstention by construction). 
We consider the machine-reported NPE and sanitizer bugs, whose bug reports always have a stack trace and reproduction test, respectively.

\begin{figure}
\includegraphics[width=0.9\columnwidth]{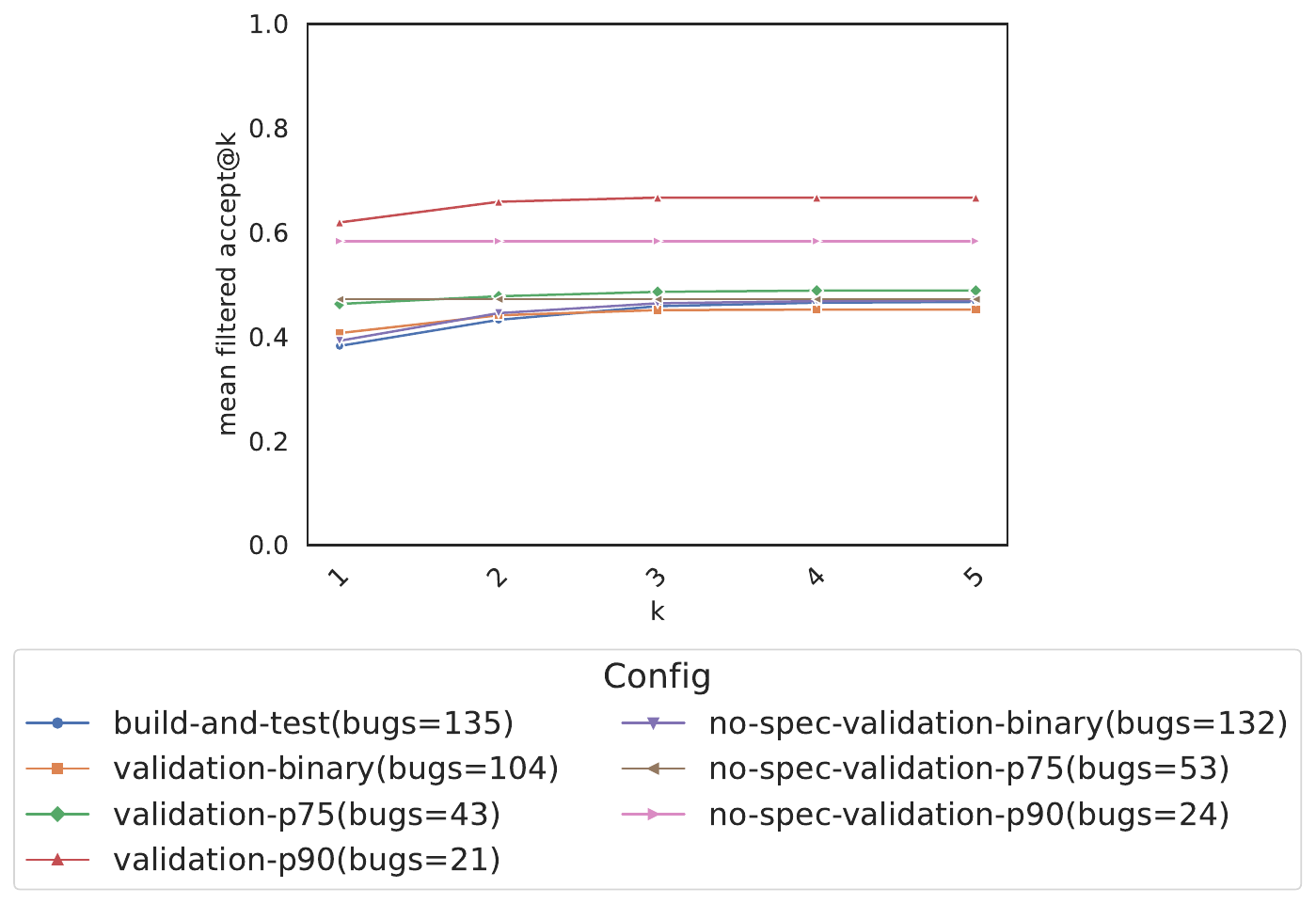}
\caption{More aggressive validation can increase filtered accept@k for NPE bugs.}
\label{fig:rq4-npe}
\end{figure}

Figure~\ref{fig:rq4-npe} shows filtered accept@k by manually inspecting the correctness of patches for NPE bugs, whose bug reports have stack traces.
The baseline (\CodeIn{build-and-test}) filters patches that did not pass build and test regression checks.
We observe that validation using binary judgment provide little improvement over these regression checks.
However, validation using the more-aggressive percentile-threshold judgment can result in improvements to filtered accept@1 of +8 percentage points and +24 percentage points, with 75th and 90th percentile thresholds, respectively.
No-spec validation at 90th percentile produces a bigger filtered accept@1 improvement than that at 75th percentile (+20 percentage points vs. +24 percentage points relative to regression checks), but both of them perform worse than the spec-based variant at 90th percentile.

\begin{figure}
\includegraphics[width=0.8\columnwidth]{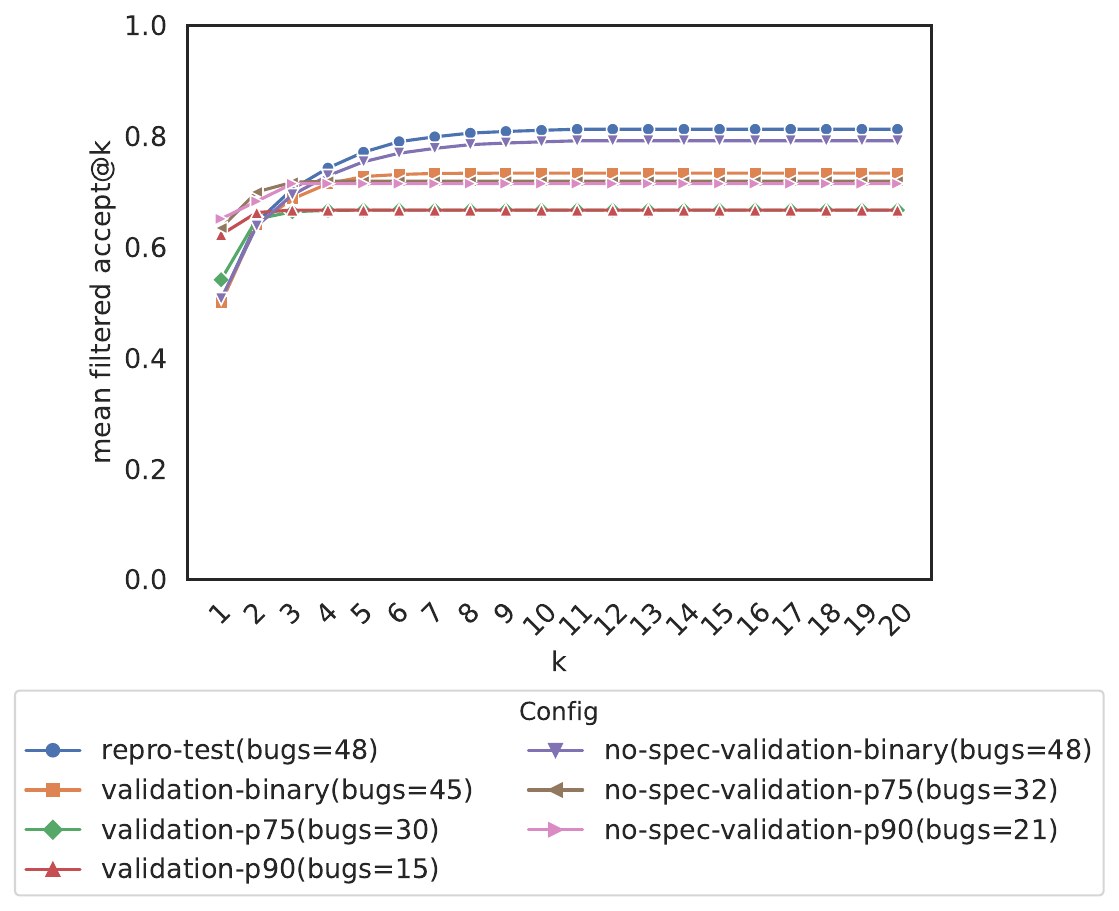}
\caption{Aggressive validation increases filtered accept@1 for sanitizer bugs. But just reviewing more patches can lead to higher filtered accept@k, with 4 patches sufficing.}
\label{fig:rq4-sanitizer}
\end{figure}

Figure~\ref{fig:rq4-sanitizer} shows mean filtered accept@k (based on manual inspection of patch correctness) for sanitizer bugs, whose bug reports have reproduction tests.
As a result, we filter any patches first by the reproduction test and then apply the validation policy. 
The baseline (\CodeIn{repro-test}) filters patches that did not pass reproduction tests.
For sanitizer bugs, validation can lift filtered accept@1 but increasing $k$ quickly outstrips the validation policy.
At $k=4$, sampling patches from the set that passes the reproduction test provides better results (0.74) than using the validation’s filtering.
When we consider $k=1$ and using 90th percentile threshold, no-spec variant \CodeIn{no-spec-validation-p90} (0.65) outperforms the spec-based variant \CodeIn{validation-p90} (0.62) and baseline \CodeIn{repro-test} (0.5).

We inspected whether patch validation performs differently based on the sanitizer bug type.
Table~\ref{tab:rq4-sanitizer-type} shows the lift for each sanitizer bug category, defined as the ratio of filtered accept@1 to the raw accept@1, where per-bug accept@1 is averaged by sanitizer type.
If a policy removes all patches associated with a sanitizer bug type or there are no patches (in the unfiltered set) that are correct, we enter NA in the table.
We find that patch validation performance indeed varies by bug type.
For the two main categories of bugs in our evaluation set—data races and use of uninitialized values—validation does not provide a meaningful improvement over just filtering with reproduction tests.
For use of uninitialized value errors, adding the binary validation and 90th percentile validation to reproduction test filtering slightly reduces lift, while adding the 70th percentile validation keeps the same lift.
For data races, adding validation actually reduces lift compared to just using reproduction tests.
We hypothesize that, for these bugs, a patch that adds a locking mechanism indeed resolves the data race and can pass patch validation, but still may not be an acceptable solution for a Google developer if it introduces downsides such as larger critical sections.
For some categories, e.g. leaks and misaligned pointer usage, a less aggressive validation achieves better lift than more aggressive ones. 

\begin{table}[]
\resizebox{\columnwidth}{!}{
\begin{tabular}{@{}cccccc@{}}
\toprule
\textbf{Sanitizer} & \textbf{\# Bugs} & \textbf{\CodeIn{repro-test}} & \textbf{\CodeIn{validation-binary}} & \textbf{\CodeIn{validation-p75}} & \textbf{\CodeIn{validation-p90}} \\ \midrule
data\_race                    & 17      & 3.11       & 2.93              & 2.43           & 3.04           \\
Fuzz+property testing   & 3       & NA         & NA                & NA             & NA             \\
heap\_buffer\_overflow          & 1       & 1.43       & 1.43              & 1.54           & 1.54           \\
invalid\_bool\_load           & 2       & 1.49       & 1.65              & 1.25           & NA             \\
invalid\_enum\_load           & 1       & NA         & NA                & NA             & NA             \\
leak\_detected                & 2       & 1.52       & 1.72              & 4.29           & 3.81           \\
misaligned\_pointer\_use      & 6       & 1.34       & 1.33              & 2.14           & 1.43           \\
null\_pointer\_use            & 1       & 2.86       & 3.33              & NA             & NA             \\
signed\_integer\_overflow     & 2       & 1.2        & 1.27              & 1.56           & 2.5            \\
stack\_use\_after\_scope      & 2       & 1.85       & 1.85              & 1.67           & 3.33           \\
use\_of\_uninitialized\_value & 13      & 2.05       & 1.93              & 2.05           & 1.95           \\ \bottomrule
\end{tabular}}
\caption{
Lift (filtered accept@1 / original accept@1) varies for different sanitizer bug types. The higher the better.
}
\label{tab:rq4-sanitizer-type}
\vspace{-10pt}
\end{table}

We hypothesize that a true bug reproduction test, such as that included in the sanitizer’s bug report, results in a set of plausible patches that are increasingly difficult for validation to assess, compared to the patches produced for the less-constrained bugs such as the human-reported ones.
This points to a few future research directions.
First, generating additional executable tests to help assess patch correctness is a particularly promising direction.
For example, AutoCodeSherpa~\cite{kang2025autocodesherpa} generates property-based tests to assist reviewing and filtering agent-generated patches.
As pointed out, after filtering with the reproduction test, sampling $k=4$ produces a mean filtered accept rate of 0.74.
Supporting developers when inspecting more patches may help maintain a high success rate.

%% file: discussion.tex
\section{Discussion}
\label{sec:discussion}

\begin{figure}[t!]
    \centering

    \begin{subfigure}[b]{0.5\columnwidth} %
        \centering
        \includegraphics[width=1.1\columnwidth]{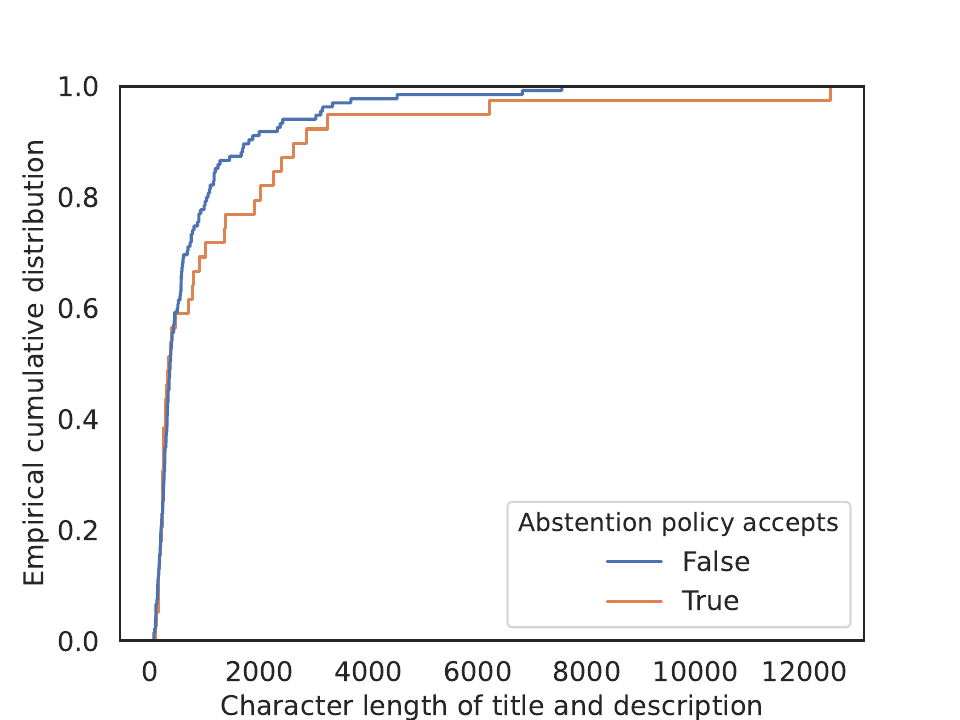}
        \caption{Bug report length}
    \end{subfigure}
~
    \begin{subfigure}[b]{0.5\columnwidth} %
        \centering
        \includegraphics[width=1.1\columnwidth]{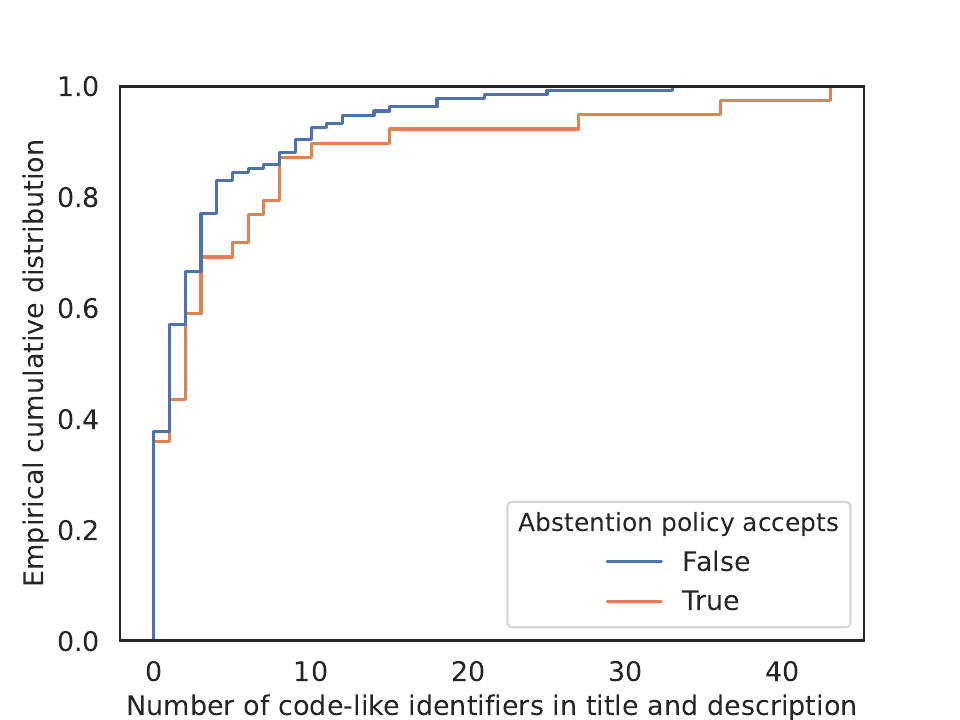}
        \caption{Code-like identifiers}
    \end{subfigure}

    \caption{Abstention-accepted bugs tend to have slightly longer title and descriptions and more code identifiers.}
\label{fig:discussion-abstention-properties}
\end{figure}

Because the abstention policy has access to the bug report when making a judgment, wee compute basic properties on our human-reported bug set to characterize their influence on policy outcomes.
As shown in Figure~\ref{fig:discussion-abstention-properties} and Table~\ref{tab:discussion-abstention-properties}, bugs accepted by abstention tend to have lengthier a bug report, and higher count of words that are possible code identifiers (based on underscore or capitalization changes).  Similarly, accepted bugs tend to have a slightly-higher presence of Google’s source repository root directory and possible code Markdown, indicated by triple backticks.

\begin{table}[]
\resizebox{0.8\columnwidth}{!}{
\begin{tabular}{@{}lrr@{}}
\toprule
\textbf{Metric} & \textbf{\CodeIn{Abstain}} & \textbf{\CodeIn{Attempt Repair}} \\
\midrule
Average \# code-like identifiers & 3.03 & 5.28 \\
Average character length & 771.67 & 1239.26 \\
Fraction with repository root & 0.12 & 0.31 \\
Fraction possible build rule & 0.01 & 0.03 \\
Fraction Markdown code ticks & 0.17 & 0.20 \\
\bottomrule
\end{tabular}
}
\caption{Abstention-accepted bugs more frequently have the presence of the monorepo root directory.}
\label{tab:discussion-abstention-properties}
\vspace{-10pt}
\end{table}

Similarly, we considered the size of the patches accepted by the 75th percentile-based patch validation for our human-reported bug set. We only consider patches that underwent LLM policy judgment. We count the number of line-level operations (addition/deletions) per patch. We find that validation-accepted patches had on average 18.38 operations, compared to 22.69 for rejected patches.

\begin{figure}[t!]
    \centering

    \begin{subfigure}[b]{0.4\textwidth}
        \centering
        \includegraphics[width=\textwidth]{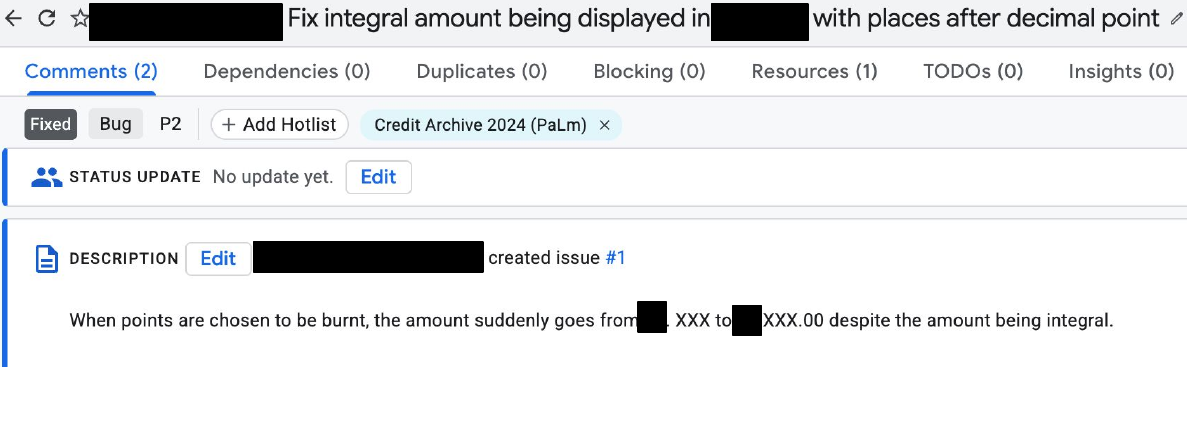}
        \caption{Description}
    \end{subfigure}

    \begin{subfigure}[b]{0.4\textwidth}
        \centering
        \includegraphics[width=\textwidth]{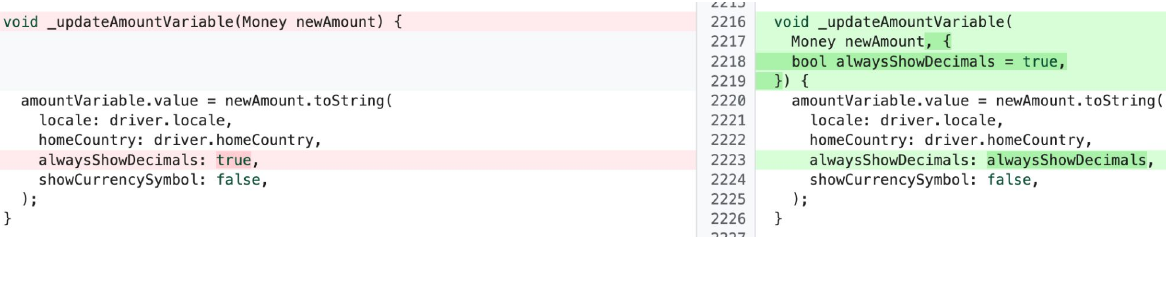}
        \caption{Ground-truth}
    \end{subfigure}

    \begin{subfigure}[b]{0.4\textwidth}
        \centering
        \includegraphics[width=\textwidth]{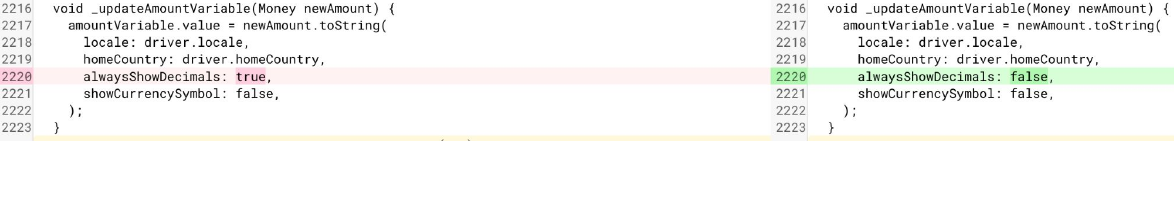}
        \caption{Agent-generated candidate patch}
    \end{subfigure}

    \caption{A bug report indicates 
    a case-dependent change. The validation's generated specification does not make this distinction so the policy incorrectly accepts the candidate.}
\label{fig:discussion-falsepositive-bug1}
\end{figure}

We now look at an example
of a patch that validation (at 90th percentile) 
incorrectly accepted.
Figure~\ref{fig:discussion-falsepositive-bug1} shows the associated report indicating cases for a change in number formatting
(when \emph{``points are chosen to be burnt''}). The report
leaves unspecified what should
happen in other cases, but a human
would likely take this to mean that other cases should still render the decimal point. 

The ground-truth patch makes decimal rendering
a function argument and modifies necessary call sites (elided for brevity). In contrast, the agent
patch modifies the constant \CodeIn{true} to \CodeIn{false}, which will impact all call sites.
The generated specification identifies the correct
method to change, but does not scope to particular call sites. The validation policy then incorrectly accepts.

%% file: threats.tex
\section{Threats to Validity}

Human and sanitizer-reported bugs were drawn from a set that has a known fix, and some of our benchmark curation relies on ground-truth patch properties (e.g., presence of a test). These constraints allow us to focus on patches within scope for an agent and use ground-truth tests for evaluation. For sanitizer-reported bugs, where reproduction tests are available at reporting time, having the ground-truth patch allows us to perform a manual comparison to determine correctness. This curation closely follows practices in open-source benchmarks like SWE-Bench and industry~\cite{maddila2025agentic, rondon2025evaluating}. 

The best-performing abstention policy includes human-authored guidelines. 
These guidelines specify high-level properties of the bug report which we believe to be general across 
application domain and organization.
Nonetheless, it is possible that such guidelines must be adapted to maintain performance in other settings.

Manual correctness assessment may diverge from the decisions that would have been made by real code owners for patches that they did not inspect. To mitigate this risk, we employed multiple annotators, consulting others in cases of ambiguity. 

Like other large-scale industrial software engineering studies~\cite{bird2011don, marginean2019sapfix, guo2010characterizing}, the bugs and code in this work are drawn from a single company (Google) and reflect internal practices. Projects or companies that write substantially different bug reports, or adhere to different development practices, may observe different policy impacts.

We carried out experiments using Gemini~2.5~Pro and 2.0~Flash.
As new LLMs become available, we expect policy performance to vary (and improve) compared to what we report here.

%% file: related.tex
\section{Related Work}

We use our existing agentic APR system Passerine~\cite{rondon2025evaluating} for evaluation. 
SpecRover~\cite{specrover} produces function-level specifications to guide the repair agent.
Similarly, our validation policy generates specifications to guide the patch judgment but these are at the file level, can be derived from multiple files, and are used only after the repair agent has concluded.
\citet{pan2025training} showed that a fine-tuned LLM can 
choose the best patch per bug based on the agent's trajectory.
Our no-spec validation employs a similar concatenation of trajectory prompt messages but uses only prompting and
filters patches rather than pick a single one.
Our validation policy also employs regression checks, or a reproduction test if available. Similarly, Agentless~\cite{xia2024agentless} uses (generated) tests for patch filtering.

Abstention~\cite{chow1970}  has long been employed in classification to reduce mispredictions due to ambiguity or lack of information. 
Past work has explored this topic extensively, learning abstention policies in boosting algorithms~\cite{cortes2016boosting}, with fairness guarantees~\cite{pmlr-v161-schreuder21a}, and, more recently, in the context of generative models~\cite{wen-etal-2025-know}.

Concurrent to our work,
BouncerBench~\cite{bouncerbench} introduces the concept of ``bouncers''
to reduce the overconfidence of an agentic software engineer. Bouncers reject underspecified tasks and unlikely patches,
closely matching our abstention and validation policies.
The benchmark BouncerBench opens up this challenge to the community and represents a complementary analysis to our own, which focuses on evaluating these policies in an industrial context.

\citet{maddila2025agentic}
concurrently introduced a patch critic in an industrial agentic APR system. We show abstention is complementary, reducing developer noise further. We also cover additional bug categories (with and without known reproduction tests).

LLM-as-judges for code~\cite{tong-zhang-2024-codejudge, crupi} are closely related to our patch validation policy, which is
analogous to reference-free judging.
However, we focus exclusively on program repair and use judgments to filter patches to reduce noise (at the expense of recall).

%% file: conclusion.tex
\section{Conclusion}
\label{sec:conclusion}

We present a dual-policy framework to reduce developer noise in agentic APR.
A bug abstention policy rejects bugs unlikely to be solved by the APR system, while a patch validation policy rejects unlikely candidate patches.
We evaluate these policies on three sets of bugs from Google’s codebase with patches produced by our agentic APR system.
On human-reported bugs, abstention and validation can raise success rates up to 13 and 15 percentage points, respectively, and up to 39 percentage points in combination.

In practical terms, this improvement transforms the developer experience: instead of expecting a successful patch roughly 1 in 10 times, developers can now expect a correct patch in 1 in 2 cases presented for their review (\CodeIn{combined-p90}). A more relaxed policy configuration (\CodeIn{combined-p75}) doubles the number of patches for inspection while still yielding a high success rate of roughly 1 in 3.

The patch validation policy can also be used to improve success rates for null pointer exceptions and sanitizer-reported bugs.

%% file: appendix.tex
\section{Human-Reported Bug Curation}
\label{app:human-bug-curation}

In this section, we describe the criteria for curating human-reported bugs that have a known ground truth patch for completeness.
The criteria closely follows \cite{rondon2025evaluating}, with the addition
of the test target criteria. Namely, each bug must satisfy the following:

\begin{enumerate}
    \item Was reported during a period of a year and closed with code changes after a model-based cutoff date in the recent past – we use the cutoff to prevent old code changes that are more likely to be stale.

    \item Has a known ground-truth fix and an associated fail-to-pass test that could be used as a held-out test oracle. This fail-to-pass test must execute without any build errors and fail in the pre-patch code and execute and succeed in the post-patch code.

    \item The bug and submitted code changes have a 1:1 relationship, mitigating risk of tangled commits

    \item Did not have screenshots in their description (multimodal repairs are out of scope for this work)

    \item Affected files for any of the following languages: C++, C, Dart, Java, Kotlin, Typescript, Javascript, Go, and Python and conversely excluded changes that affected SQL, HTML, CSS, configuration languages, binary data files, or multimedia data files

    \item Had a unidiff ground-truth commit of less than 150 lines

    \item Had <= 31 test targets that depended on the files changed in the ground-truth patch.
\end{enumerate}

\section{Bug Abstention Prompts}
\label{app:abstention-prompts}

The section contains the full prompts for the bug abstention policies described in Section~\ref{sec:abstention-bug-representations}.

\begin{lstlisting}[basicstyle=\footnotesize\ttfamily, caption=Prompt for bug representation $r_1(b)$., label=lst:abstention-r1-prompt]
You are tasked to predict whether an LLM-based agentic approach will successfully fix a bug.
The agent can use a suite of tools by issuing commands, including:
    - `code_search` to find relevant code.
    - `cat` to read file contents.
    - `ls` to list directory contents.
    - `find_and_replace_text` to modify files.
    - `bazel` to run tests.
The agent is presented with a bug report and can perform a maximum of 25 steps.

Analyze the following bug and predict if the agent will succeed. Your output must be a single word: either "success" or "failure". Do not provide any other text, explanation, or punctuation.

# Bug Title
{bug.title}

# Bug Description
{bug.description}
\end{lstlisting}

\begin{lstlisting}[basicstyle=\footnotesize\ttfamily, caption=Prompt for bug representation $r_2(b)$., label=lst:abstention-r2-prompt]
# GUIDELINES
 These guidelines provide criteria for determining if a bug is likely to be solved by the agent.
 - Clear Problem & Action: The bug report explicitly and clearly describes the incorrect behavior, and the required change is well-understood.
 - Precise Code Localization: The report points to the exact file, code line, function, or configuration that needs attention, or it's easily identifiable from provided information (e.g., stack traces).
 - Simple, Localized Fix Anticipated: The likely solution involves a small, straightforward, and localized change.
 - Actionable Reproduction & Sufficient Context: The issue can be easily and reliably reproduced with the information provided, and sufficient context is available within the report for the agent to proceed effectively.
 - Self-Contained and Complete: All necessary information for addressing the bug is provided and self-contained within the report, requiring minimal external look-up or inference.
\end{lstlisting}

\section{Patch Validation Prompts}

This section contains the full prompts for the patch validation policy described in Section~\ref{sec:validation}.
Listing~\ref{lst:validation-specification-prompt} shows the prompt for the specification generation step.

\begin{lstlisting}[basicstyle=\footnotesize\ttfamily, caption=Using the original bug description and file contents to generate a specification., label=lst:validation-specification-prompt]
SYSTEM:
You are an expert software engineer that reviews a bug and lists out a specification for the desired behavior.
You should cover properties desirable for a fix. Be specific, do not provide general properties that apply to all bugs. Do not provide properties that are about artifacts like documentation.

USER:
Here is a bug, please create a fix specification:

# Bug Title
{bug.title}

# Bug Description
{bug.description}

# Original source code
$ cat {file_path}
{file_content}

# Specification
\end{lstlisting}

Listing~\ref{lst:validation-review-prompt} shows
the LLM prompt used for validating a patch, along
with the use of the generated specification.

\begin{lstlisting}[basicstyle=\footnotesize\ttfamily, caption={Using the original bug report, candidate patch, any test information, and the fix specification generated in the prior step, to decide if a patch is likely to fix a bug.}, label=lst:validation-review-prompt]
SYSTEM:
SETTING: You are a software engineer reviewing code changes to fix bugs. Your job is to review how a bug2code agent tried to fix a bug.
Return a ReviewResponse with whether the bug is fixed and explaining why.

USER:
Here's a log of an attempt by a bug2code agent to fix a bug:
# Bug Title
{bug.title}

# Bug Description
{bug.description}

# Patch
{patch}

# Test information
{test_information}

# Specification:
Based on the bug report and source code, here is a specification for the desired behavior of the fix.

### Fix Specification
{specification}
\end{lstlisting}

\subsection{Additional Results}

Table~\ref{tab:rq2-pr-binary} presents the per-label precision/recall for binary validation policies.

\begin{table}[]
\begin{tabular}{@{}cccc@{}}
\toprule
Approach                    & Binary Label & Precision & Recall \\ \midrule
Prevalence (build/test)     & TRUE         & 0.19      & -      \\
No-spec validation (binary) & FALSE        & 0.99      & 0.07   \\
No-spec validation (binary) & TRUE         & 0.2       & 1      \\
Validation (binary)         & FALSE        & 0.91      & 0.58   \\
Validation (binary)         & TRUE         & 0.3       & 0.76   \\ \bottomrule
\end{tabular}
\caption{Precision/Recall Table on the binary judgment: validation’s binary judgment provides modest predictive performance. Removing the specification usage reduces performance substantially.}
\label{tab:rq2-pr-binary}
\end{table}

To focus on the validation’s ability to identify successful trajectories, we also consider a variant of recall. Specifically, we compute the average (over bugs) probability that at least one trajectory is accepted by the validation, given a sample of k trajectories that satisfy the held-out reproduction test. In effect, if our repair system produced only good fixes, how often would we show at least one repair to a user if we generated k trajectories. We denote this mean (validation recall)@k. To compute percentile-based thresholding we determine cutoffs over all trajectories that pass build and test regressions first, before filtering.

\begin{figure}
\includegraphics[width=\columnwidth]{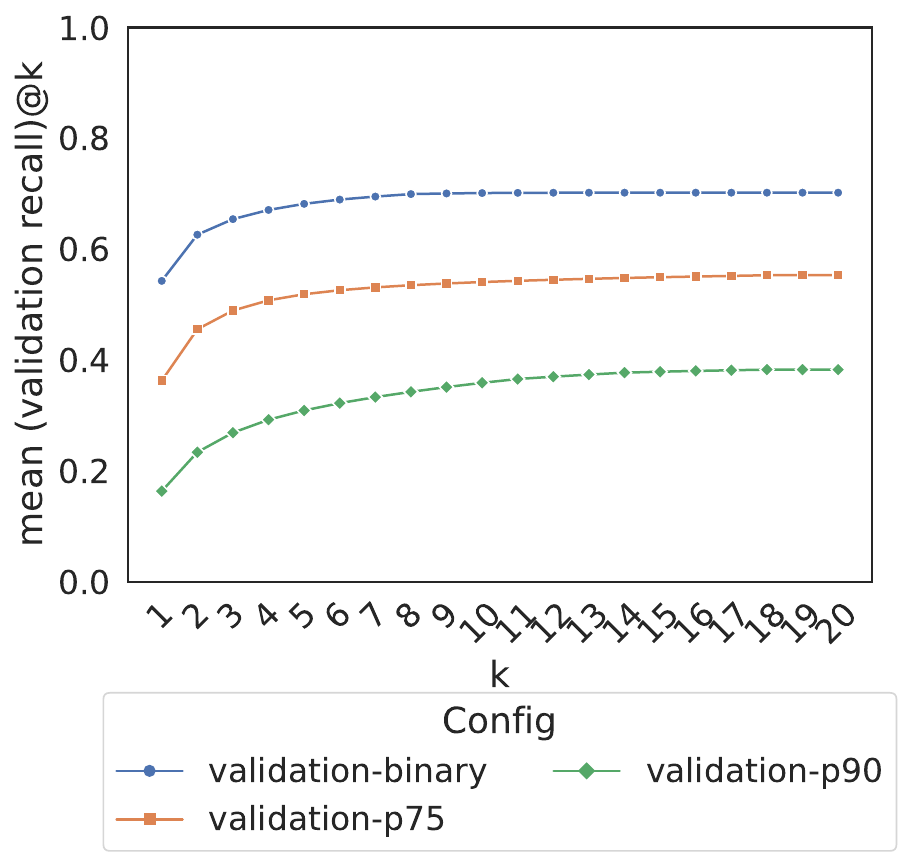}
\caption{If we consider samples of size k of trajectories that pass the held-out reproduction test, more aggressive validation is more likely to incorrectly reject patches.
}
\label{fig:rq2-pass-and-critic-recall}
\end{figure}

As shown in Figure~\ref{fig:rq2-pass-and-critic-recall},  percentile-based thresholding results in more false negatives – rejecting patches that we should have shown to humans. When sampling a single successful patch, the binary validation will show a patch at a rate of 0.54, compared to 0.36 and 0.16 for 75th and 90th percentile thresholding, respectively.

\begin{figure}
\includegraphics[width=\columnwidth]{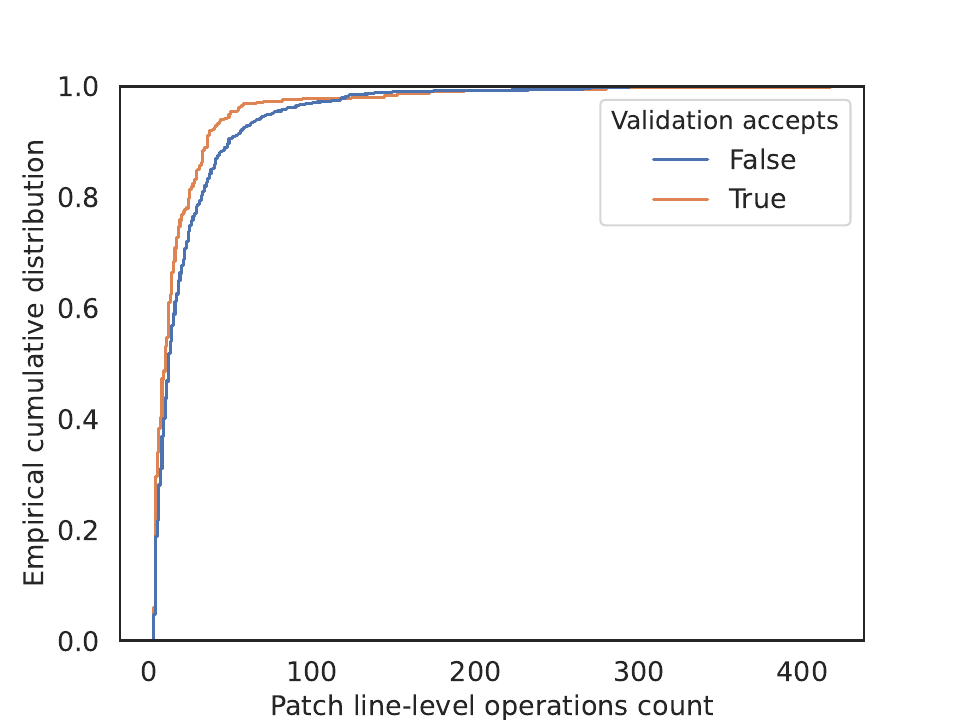}
\caption{Patches accepted by the patch validation policy have slightly fewer line-level operations compared to those it rejected.}
\label{fig:discussion-validation-properties}
\end{figure}

Figure~\ref{fig:discussion-validation-properties} shows
that patches accepted by the validation policy have
slightly more line-level operations compared to those
rejected.

\subsection{Patch Validation Decisions: False Positives}

Listing~\ref{lst:discussion-example-spec} shows the generated specification
and validation policy output for the false positive example discussed in 
Section~\ref{sec:discussion}.

\begin{lstlisting}[basicstyle=\footnotesize\ttfamily, label=lst:discussion-example-spec]
"Here is a fix specification for the bug.

### Desired Behavior Specification

1.  **Conditional Decimal Formatting for Amounts**
    The amount displayed in the amount input field must be formatted based on whether it is an integral value.
    *   If the `Money` object represents an integer (e.g., the fractional part is zero), it must be formatted as a string without decimal places (e.g., "500").
    *   If the `Money` object represents a non-integer value (e.g., the fractional part is non-zero), it must be formatted with the standard two decimal places (e.g., "500.50").

2.  **Consistent Formatting on Reward Application**
    When a user applies <redacted> points (or any other deduction) that results in an integral final amount, the amount field must update to show the new integral amount without adding trailing decimals. For example, if the original amount is "500" and points worth "10" are applied, the field must display "490", not "490.00".

3.  **Consistent Formatting on Reward Removal**
    When the user deselects the applied points and the amount reverts to its original integral value, the display must also revert to the format without decimals. For example, if the amount with points was "490" and the points are removed, the field must revert to displaying "500".

4.  **Scope of Formatting Change**
    The conditional decimal formatting logic should be applied in the `_updateAmountVariable` method. This ensures that any programmatic update to the amount field, including the initial amount load, application of charges, and application/removal of deductions, follows the same consistent formatting rule

----

is_bug_fixed: true
explanation: "The patch correctly fixes the bug. The bug description states that integral amounts are being displayed with unnecessary decimal points (e.g., \"XXX.00\") after rewards are applied. The fix specification requires conditional formatting: show decimals only for non-integer amounts. The patch modifies the `_updateAmountVariable` method by changing the `alwaysShowDecimals` parameter from `true` to `false` in the `newAmount.toString()` call. This change ensures that the `Money` object\'s string representation will only include decimal places if the amount is not an integer, which directly addresses the root cause of the bug and fulfills all requirements of the specification."
internal_score: 0.62615377
\end{lstlisting}

\begin{figure}[ht]
    \centering

    \begin{subfigure}[b]{\columnwidth}
        \centering
        \includegraphics[width=\columnwidth]{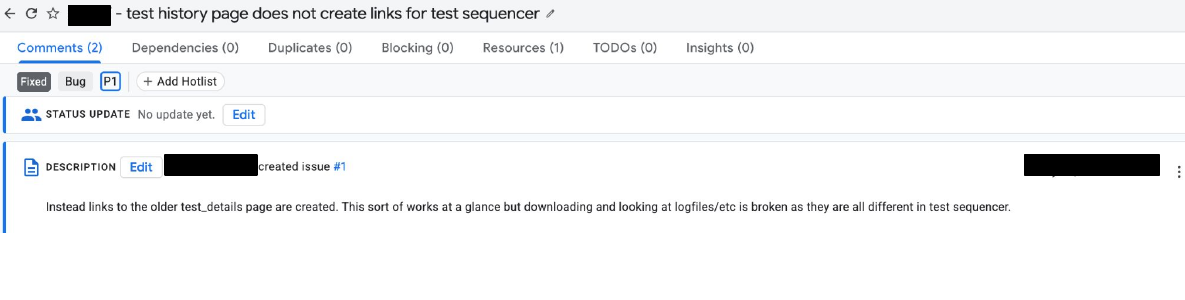}
        \caption{Description}
    \end{subfigure}

    \begin{subfigure}[b]{\columnwidth}
        \centering
        \includegraphics[width=\columnwidth]{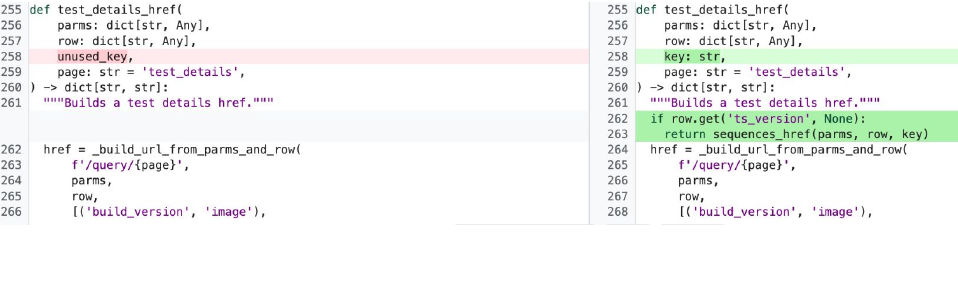}
        \caption{Ground-truth}
    \end{subfigure}

    \begin{subfigure}[b]{\columnwidth}
        \centering
        \includegraphics[width=\columnwidth]{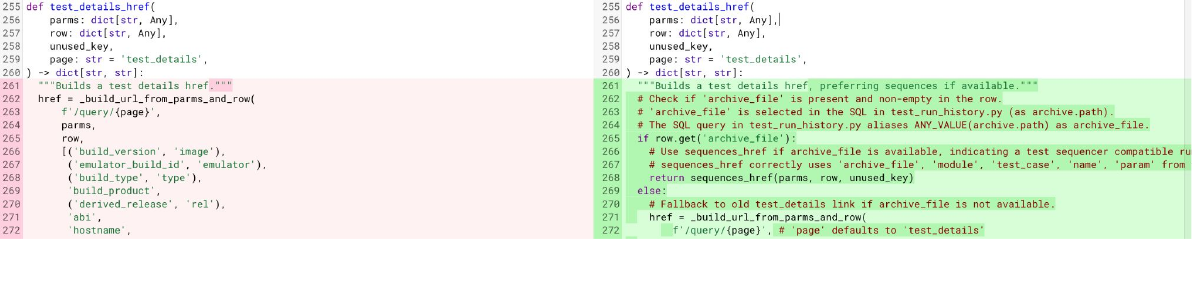}
        \caption{Agent Patch}
    \end{subfigure}

    \caption{A false positive in patch validation.}
\label{fig:discussion-falsepositive-bug2}
\end{figure}

In the next case (Figure~\ref{fig:discussion-falsepositive-bug2}), the bug report provides relatively high-level discussion of the issue but does not specify what property should be used to check for links for a test sequencer module.

The ground-truth patch adds conditional handling based on the presence of a particular field (ts\_version) in an input row. The agent’s patch modifies this into an if-else statement. The else branch is equivalent to the fall-through case in the ground-truth patch. However, the if guard checks a different field (archive\_file), which it inferred from the existing test files.

The generated specification correctly points out that links for a different module will need to be generated in a conditional fashion. It also elaborates on where these changes should be made and what field to check in the input. The validation’s output correctly reflects the specification. However, the specification was wrong to begin with, as the field to be checked in the input is actually different: the field that the agent inferred was ultimately incorrect. In cases such as these, it is likely that a small piece of human feedback on the final patch would allow either the validation policy to make the right decision (reject) or the repair agent to update the patch.

\begin{lstlisting}[basicstyle=\footnotesize\ttfamily]
Here is a specification for the fix.

### Fix Specification

The root cause of the bug is that the `test_details_href` function unconditionally generates links to the `test_details` page, even for test runs originating from the "Test Sequencer", which require a different details page (`/query/sequences`).

The fix will involve modifying the `test_details_href` function to correctly identify Test Sequencer runs and generate the appropriate URL.

1.  **Conditional Link Generation in `test_details_href`**
    *   The function `test_details_href` in `url_builder.py` must be modified to conditionally generate one of two different URLs based on the properties of the test run.

2.  **Identification of Test Sequencer Runs**
    *   A test run shall be identified as a "Test Sequencer" run if the `row` dictionary passed to `test_details_href` contains the key `'archive_file'`. The presence of this key is the designated indicator for a sequencer test result.

3.  **URL Generation for Test Sequencer Runs**
    *   If a test run is identified as a Test Sequencer run (i.e., `'archive_file' in row` is true), the `test_details_href` function must delegate the URL creation to the existing `sequences_href` function.
    *   The `parms` and `row` arguments received by `test_details_href` must be passed directly to `sequences_href` to ensure all necessary parameters (`archive_file`, `module`, `test_case`, `name`, `param`) are available for building the correct `/query/sequences` URL.

4.  **Preservation of Existing Behavior for Non-Sequencer Runs**
    *   If a test run is not identified as a Test Sequencer run (i.e., `'archive_file'` is not in `row`), the function must retain its original behavior.
    *   It must proceed to build a URL pointing to `/query/test_details` (or the page specified by the optional `page` parameter) using the existing `_build_url_from_parms_and_row` call and its original list of parameters.

----

is_bug_fixed: true
explanation: "The patch correctly addresses the bug by modifying the `test_details_href` function. It now checks if the `row` data contains an `archive_file` key, which is the designated indicator for a test sequencer run. If the key is present, it correctly calls the `sequences_href` function to generate the appropriate link. If not, it maintains the original behavior, generating a link to the older `test_details` page. This conditional logic perfectly matches the fix specification and resolves the issue of incorrect link generation."
internal_score: 0.59506536
\end{lstlisting}

\subsection{Patch Validation Decisions: True Positives}

For the two bugs below, the agent patch is able to pass the held-out ground-truth reproduction test. Additionally, manual inspection shows these are equivalent to the changes made in the ground-truth patch.

\begin{figure}[ht]
    \centering

    \begin{subfigure}[b]{\columnwidth}
        \centering
        \includegraphics[width=\columnwidth]{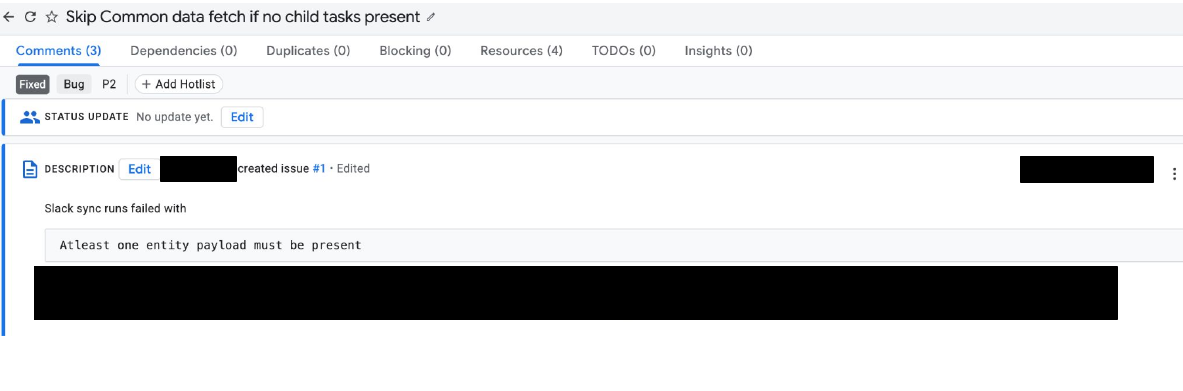}
        \caption{Description}
    \end{subfigure}

    \begin{subfigure}[b]{\columnwidth}
        \centering
        \includegraphics[width=\columnwidth]{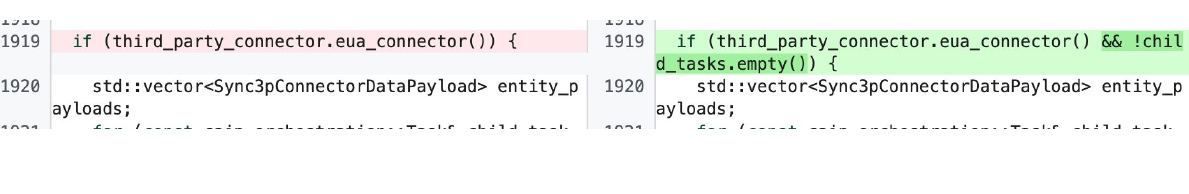}
        \caption{Groundtruth}
    \end{subfigure}

    \begin{subfigure}[b]{\columnwidth}
        \centering
        \includegraphics[width=\columnwidth]{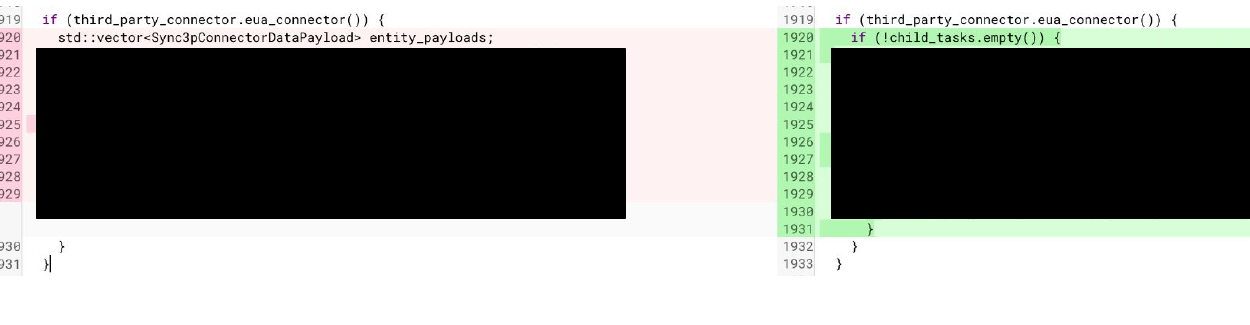}
        \caption{Agent Patch}
    \end{subfigure}

    \caption{A true positive in patch validation.}
\label{fig:discussion-truepositive-bug1}
\end{figure}

The first bug (Figure~\ref{fig:discussion-truepositive-bug1}) requires that the agent add an emptiness check to avoid processing cases where there are no child tasks. The ground-truth patch tightens a guard to add this check.

The specification generation step clearly outlines what the changes necessary should be and the review explanation shows that the agent patch adheres to that specification and so is judged to be correct.

\begin{lstlisting}[basicstyle=\footnotesize\ttfamily]
"Based on the bug report, here is a specification for the fix.

### Desired Behavior Specification

1.  **Conditional Common Data Fetch Execution:**
    *   In the `<redacted>` function, the call to `ExecuteCommonDataFetchUsingEua` must only be performed if the list of generated `child_tasks` is not empty.

2.  **Graceful Handling of No-Op Syncs:**
    *   If `CreateChildTasks` returns an empty vector of `child_tasks` (for example, because all entities for the connector are already in a `RUNNING` state), the `HandleParent` function must skip the common data fetch logic.
    *   The function should then proceed to return a `<redacted>::Success` with an empty list of tasks. This indicates that the parent task has completed successfully without scheduling any new work.

3.  **Preservation of Existing Validations:**
    *   The argument validation inside `ExecuteCommonDataFetchUsingEua` which checks if the `entity_payloads` vector is empty and returns an `InvalidArgumentError` must not be removed. The fix must prevent this function from being called with an empty vector, rather than altering the function's internal guards.

----
is_bug_fixed: true
explanation: "The bug occurs when the common data fetch logic (`ExecuteCommonDataFetchUsingEua`) is called with an empty list of child tasks, leading to an \"Atleast one entity payload must be present\" error. The patch correctly fixes this by wrapping the entire common data fetch logic, including the unpacking of payloads and the call to `ExecuteCommonDataFetchUsingEua`, inside a conditional block `if (!child_tasks.empty())`. This ensures that the data fetch is only attempted when there are actual child tasks to process, preventing the error and allowing the parent task to complete successfully with no new tasks scheduled, which is the correct behavior for a no-op sync."
internal_score: 0.65
\end{lstlisting}

\begin{figure}[ht]
    \centering

    \begin{subfigure}[b]{\columnwidth}
        \centering
        \includegraphics[width=\columnwidth]{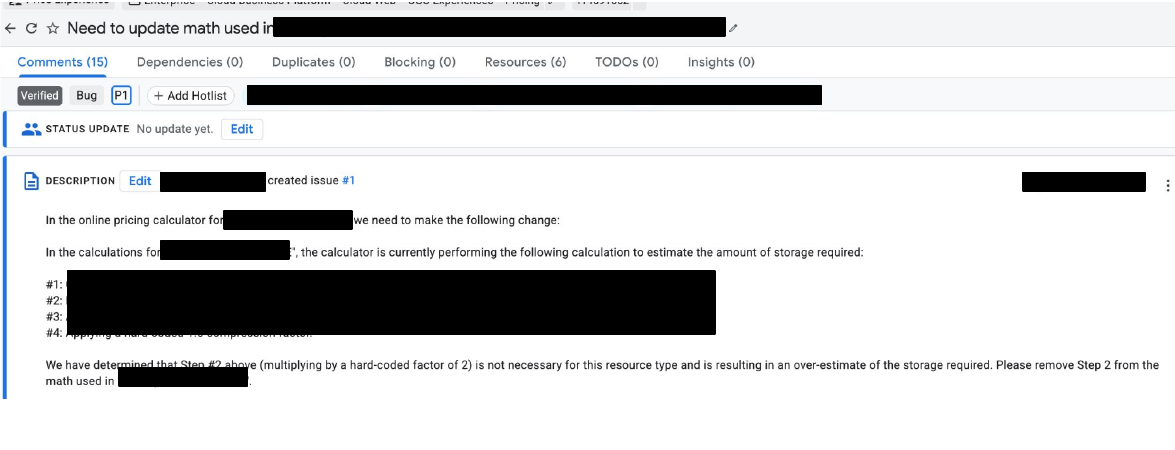}
        \caption{Description}
    \end{subfigure}

    \begin{subfigure}[b]{\columnwidth}
        \centering
        \includegraphics[width=\columnwidth]{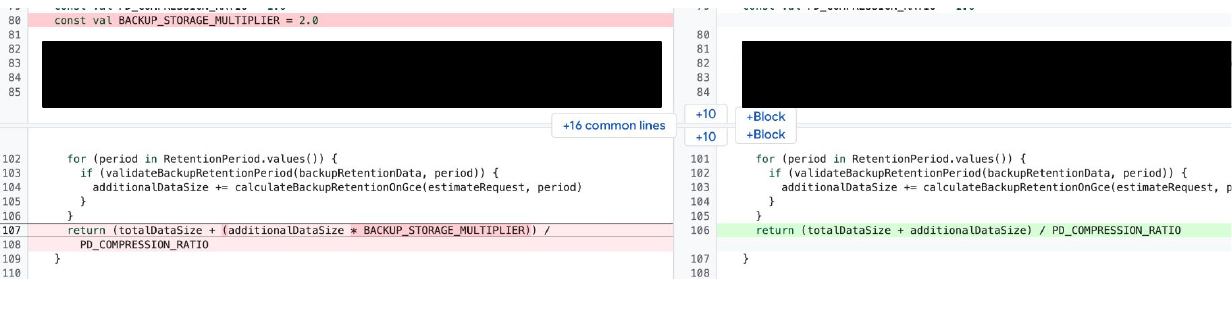}
        \caption{Ground-truth}
    \end{subfigure}

    \begin{subfigure}[b]{\columnwidth}
        \centering
        \includegraphics[width=\columnwidth]{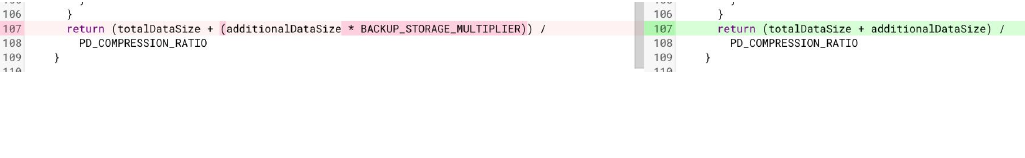}
        \caption{Agent Patch}
    \end{subfigure}

    \caption{A true positive in patch validation.}
\label{fig:discussion-truepositive-bug2}
\end{figure}

In the next example (Figure~\ref{fig:discussion-truepositive-bug2}), the bug description provides instructions on the expected edits to a calculation. The ground-truth patch updates the calculation and in addition removes a constant that is no longer used. The agent patch \emph{only} updates the calculation, but does so in the correct way.

The generated specification closely reflects the bug description, given the level of detail in that report. However, notice it has additional requirements such as limiting the scope of the changes. The generated validation output judges the patch is correct based on the specification produced.

\begin{lstlisting}[basicstyle=\footnotesize\ttfamily]
Here is a fix specification for the bug.

### **Specification for <redacted> Calculation Fix**

This specification outlines the required changes to correct the Total Cost of Ownership (TCO) calculation for the <redacted> in the pricing calculator.

#### **1. Functional Behavior**

The estimated backup storage calculation for <redacted> must be corrected to no longer multiply the incremental backup data size by a factor of 2.

-   **Current Incorrect Calculation Logic:**
    `Estimated Storage = (Total Data Size + (Incremental Data Size * 2.0)) / 1.6`

-   **Required Correct Calculation Logic:**
    `Estimated Storage = (Total Data Size + Incremental Data Size) / 1.6`

#### **2. Specific Requirements**

1.  **Removal of Redundant Multiplier:**
    In the `calculateEstimatedBackupStorage` function within the `<redacted>` file, the multiplication by the `BACKUP_STORAGE_MULTIPLIER` constant must be removed.

    -   **Location:** `<redacted>`, line 111
    -   **Current Code:** `return (totalDataSize + (additionalDataSize * BACKUP_STORAGE_MULTIPLIER)) / PD_COMPRESSION_RATIO`
    -   **Required Code:** `return (totalDataSize + additionalDataSize) / PD_COMPRESSION_RATIO`

2.  **Correct Aggregation of Storage Components:**
    The final estimated storage value must be the result of summing the initial full backup size (`totalDataSize`) with the aggregated incremental changes (`additionalDataSize`), and then dividing the sum by the `PD_COMPRESSION_RATIO`.

#### **3. Invariants (Behaviors that must not change)**

1.  **Scope of Change:** The modification must be strictly confined to the `<redacted>`. Calculations for other Backup and DR services must not be affected.
2.  **Compression Ratio:** The `PD_COMPRESSION_RATIO` (value 1.6) must continue to be applied to the total sum of the base data and all incremental data.
3.  **Incremental Data Calculation:** The logic for calculating the `additionalDataSize` (as performed by the `<redacted>` function) must remain unchanged. The fix only addresses how this value is used in the final aggregation, not how it is derived.
4.  **Base Data Inclusion:** The `totalDataSize` (representing the initial full backup) must continue to be included as the base amount of data stored before incremental changes are added.

----
 is_bug_fixed: true
explanation: "The patch correctly addresses the bug by removing the multiplication by `BACKUP_STORAGE_MULTIPLIER` in the `<redacted>` file. The bug report explicitly requested the removal of a hard-coded factor of 2 that was being applied to the incremental backup data size. The code change from `(totalDataSize + (additionalDataSize * BACKUP_STORAGE_MULTIPLIER)) / PD_COMPRESSION_RATIO` to `(totalDataSize + additionalDataSize) / PD_COMPRESSION_RATIO` directly implements this requirement, thus fixing the over-estimation in the <redacted> calculation."

\end{lstlisting}

In general, more detailed bug reports are likely to produce better bug abstention and patch validation results. For the former, the validation policy can create a more precise set of guidelines, which can then be used to judge the agent’s patches.

\subsection{Optimizability of Decisions Based on Bug Abstention and Patch Validation policies}

While in this work, we focus on reducing noise for human developers by avoiding showing them patches that are unlikely to be correct,  the decision of what bugs to tackle and what patches to surface can be approached as a more complex optimization problem. For example, we might explicitly model the effort and cost of human review, and use this to guide our choices. Below we briefly introduce some simple metrics that could be used and leave this for future work.

Let $V_b$ be the value of successfully closing a bug $b$ with a fix. Let $C_b$ be the cost on the code base of leaving the bug unrepaired. Let $R_c$ be the cost to produce a repair trajectory. Let $p^*$ be the true, unknown, probability of a trajectory solving bug $b$.

The expected net value of the abstention policy is then:

$$
\begin{cases}{ll}
p^*(V_b)+(1-p^*)(-C_b)-R_c & \text{if } P_a(b) \ge \tau_a \\
-C_b & \text{otherwise}
\end{cases}
$$

Let $H_c$ be the cost of having a human review a patch from trajectory $t$. The analogous expected net value for the patch validation policy is then:

$$
\begin{cases}{ll}
p^*(V_b)+(1-p^*)(-C_b)-H_c & \text{if } D_v(P_v(b, t)) \\
-C_b & \text{otherwise}
\end{cases}
$$